\documentclass[reprint,prx,10pt,superscriptaddress,twocolumn]{revtex4-1}

\usepackage{amsmath}    % need for subequations
\usepackage{graphicx}   % need for pdf figures
\usepackage{epstopdf}	% need for eps figures
\usepackage{verbatim}   % useful for program listings
\usepackage{color}      % use if color is used in text
\usepackage{subfigure}  % use for side-by-side figures
\usepackage{hyperref}   % use for hypertext links, including those to external documents and URLs
\newcommand{\bo}{\boldsymbol}

\begin{document}
%\begin{bibunit}[h-physrev]
\title{\textbf{Josephson Interference due to Orbital States in a Nanowire Proximity Effect Junction}}

\author{Kaveh Gharavi}
\affiliation{Institute for Quantum Computing, University of Waterloo, Waterloo, Ontario N2L 3G1, Canada}
\affiliation{Department of Physics and Astronomy, University of Waterloo, Waterloo, Ontario N2L 3G1, Canada}
\author{Gregory W. Holloway}
\affiliation{Institute for Quantum Computing, University of Waterloo, Waterloo, Ontario N2L 3G1, Canada}
\affiliation{Department of Physics and Astronomy, University of Waterloo, Waterloo, Ontario N2L 3G1, Canada}
\author{Chris M. Haapamaki}
\affiliation{Institute for Quantum Computing, University of Waterloo, Waterloo, Ontario N2L 3G1, Canada}
\author{Mohammad H. Ansari}
\affiliation{Institute for Quantum Computing, University of Waterloo, Waterloo, Ontario N2L 3G1, Canada}
\affiliation{Kavli Institute of Nanoscience, Delft University of Technology, Lorentzweg 1, 2628 CJ, Delft, the Netherlands}
\author{Mustafa Muhammad}
\affiliation{Institute for Quantum Computing, University of Waterloo, Waterloo, Ontario N2L 3G1, Canada}
\author{Ray R. LaPierre}
\affiliation{Department of Engineering Physics, Centre for Emerging Device Technologies,
McMaster University, Hamilton, Ontario L8S 4L7, Canada}
\author{Jonathan Baugh}
\email{baugh@uwaterloo.ca}
\affiliation{Institute for Quantum Computing, University of Waterloo, Waterloo, Ontario N2L 3G1, Canada}
\affiliation{Department of Physics and Astronomy, University of Waterloo, Waterloo, Ontario N2L 3G1, Canada}
\affiliation{Department of Chemistry, University of Waterloo, Waterloo, Ontario N2L 3G1, Canada}
\date{\today}

\begin{abstract}
The Josephson supercurrent in a Nb-InAs nanowire-Nb junction was studied experimentally. The nanowire goes superconducting due to the proximity effect, and can sustain a phase coherent supercurrent. An unexpected modulation of the junction critical current in an axial magnetic field is observed, which we attribute to a novel form of Josephson interference, due to the multi-band nature of the nanowire. Andreev pairs occupying states of different orbital angular momentum acquire different superconducting phases, producing oscillations of the critical current versus magnetic flux. We develop a semi-classical multi-band model that reproduces the experimental data well. While spin-orbit and Zeeman effects are predicted to produce similar behaviour, the orbital effects are dominant in the device studied here. This interplay between orbital states and magnetic field should be accounted for in the study of multi-band nanowire Josephson junctions, in particular, regarding the search for signatures of topological superconductivity in such devices.
\end{abstract}

\maketitle
There has been much recent focus on topological superconductors because of the anyonic character of the states they support \cite{nayak_review}, and their possible applications to quantum information processing \cite{nayak_review, bravyi_review}. A promising avenue for the realization of topological superconductors is via the superconducting proximity effect. Majorana fermions (MFs) \cite{maj} are predicted to be observed with a combination of conventional $s$-wave superconductivity, a proximate semiconductor with strong spin-orbit coupling and a suitable Zeeman splitting \cite{lutchynPRL2010_theory,stanescuPRB2011_theory,satoPRL2009_theory,sauPRL2010_theory,aliceaPRB2010_theory, oregPRL2010_theory}. Low bandgap semiconductor (InAs, InSb) nanowires have attracted attention as candidates for the realization of such hybrid devices. They form transparent, Schottky  barrier-free contacts with superconductors such as Nb and Al \cite {schapers_book}, have a strong Rashba spin-orbit coupling, and their high Land{\'e} $g$-factor allows them to be driven into the topological phase in the presence of a magnetic field applied perpendicular to the spin-orbit vector. Several proposals for observing MFs in such devices have been put forward \cite{lutchynPRL2010_theory, lawZBA09,flensZBA10,sauZBA10,fluxQubitReadout10,acJosephson11,qdDetection11,splitting_smoking_gun_12}, and several reports of zero-bias anomalies (ZBAs) in differential tunnelling conductance \cite{MourikSci12,Das2012,DengLundObs12} consistent with MFs \cite{PradaPRB2012} have been made. However, similar ZBAs may arise from a non-topological origin such as strong disorder \cite{tewari_review,Bagrets2012,Pientka2012,Liu2012}, Kondo resonances \cite{Lee2012}, or smooth confinement potentials in the nanowire \cite{PradaPRB2012, tewari_review,Kells2012,Rainis2013,Stanescu2013}. Indeed, there have been observations of ZBAs from apparently non-Majorana origins \cite{finkUrbanaObs13,ChangHarvardObs13,defranc2014}.\\
\indent The topological phase is theoretically predicted to arise in multiband nanowires, even in the presence of moderate disorder \cite{multibandPhysRevLett.106}. The structure of transverse subbands due to radial confinement in semiconductor nanowires has been the subject of several recent studies \cite{schapers_shell_filling_10, schapers_shell_filling_13}. Since proximity superconductivity is mediated by the Andreev reflection of electron-hole pairs \cite{prox_andreev}, and the constituent carriers occupy certain transverse subbands in the nanowire, one might expect an interplay between the proximity effect and the nanowire subband structure.  In this paper, we ask how this interplay affects the critical current of a semiconductor nanowire Josephson junction in the presence of a magnetic field. The result is crucial for properly interpreting experiments on nanowire Josephson junctions, particularly with respect to recently proposed MF detection protocols which rely on the measurement of the critical current rather than ZBAs \cite{AguadoSCreadout}. Motivated by experimental observations, we develop a model describing a novel form of Josephson interference arising in a nanowire junction under an applied axial magnetic field. The axial field orientation is needed to reach the topological phase in InAs and InSb junctions. It is shown that the interference model can explain our experimental observations on a Nb-InAs nanowire-Nb junction.
\section*{Results}
\subsection*{Nanowire-based Josephson Junction}
\begin{figure}[h!]
\centering
\includegraphics[width=0.48\textwidth]{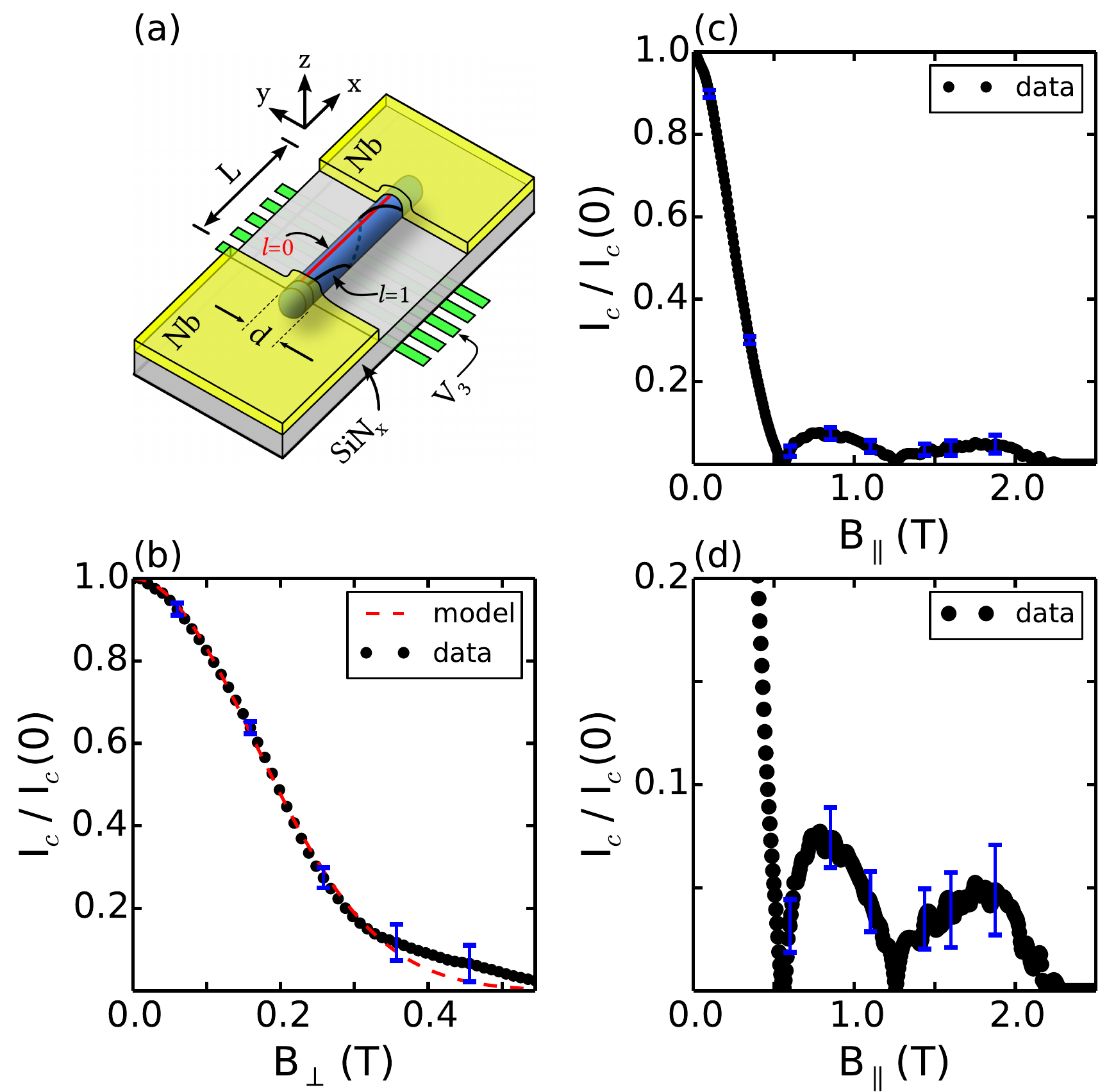}
\caption{\label{fig:sketch}
(a) Schematic of the nanowire $SNS$ junction. The set of bottom gates, especially $V_3$, are used to modulate the local potential. Our model treats the nanowire section as a cylindrical shell conductor of diameter $d$, contacted by Nb leads. We use $d = 63\; \mathrm{nm}$, from a scanning electron micrograph of the device studied experimentally. 
(b) Monotonic, quasi-Guassian decay of the junction critical current $I_c$ in a perpendicular magnetic field. The data (solid circles) can be fit to a Gaussian function, $e^ {(-0.499 (\Phi / \Phi_0)^2)}$ (dashed line), where $\Phi$ is the flux corresponding to the perpendicular field and $\Phi_0$ is the flux quantum.
(c) Oscillatory behaviour of $I_c$ in an axial magnetic field. (d) The same data as (c), enlarged to show nodes. We explain these oscillations as Josephson interference due to classical spiral trajectories of Andreev pairs on the cylindrical shell. Different paths correspond to nanowire subbands of different orbital angular momentum. Example trajectories corresponding to $l=0, 1$ are shown in (a). Representative error bars are shown for selected data points in (b,c,d).}
\end{figure}

A Superconductor/Normal conductor/Superconductor ($SNS$) Josephson junction was fabricated, wherein an InAs nanowire is used as the $N$ weak link, and is contacted by Nb leads as shown schematically in Figure \ref{fig:sketch}a. Extensive dc electrical measurements of the junction were made in a dilution refrigerator with a base lattice temperature of $25\; \mathrm{mK}$. A superconducting proximity effect is observed in the junction in the form of a dissipationless current. When the current bias exceeds a switching current value $I_{\mathrm {sw}}$, the junction switches to the normal state. The value for the switching current depends on the voltages applied to the local gates, and can be as high as $55\; \mathrm{nA}$. The phase dynamics of the junction are overdamped, so $I_{\mathrm{sw}}$ approximates the thermodynamic critical current $I_c$. Conductance modulations at voltages $V_n = 2\Delta_{\mathrm{Nb}} / (en)$, for integer $n$, signify multiple Andreev reflection and indicate phase coherence across the junction. Here, $\Delta_{\mathrm{Nb}} = 1.2\; \mathrm{meV}$ is the superconducting energy gap in the Nb leads, and $e$ is the electronic charge. An excess current \cite{otbk_corrected} of about $42\; \mathrm{nA}$  indicates a Nb-InAs contact transparency $t \sim 0.65$ -- see Supplementary Information. The figure of merit product $I_c R_N$ of the junction is $\sim 0.4 \; \mathrm{mV}$, where $R_N$ is the normal state resistance of the junction. The normal section of the junction is semiconducting, and tuning the local potential with voltages on the bottom gates, especially $V_3$ (see Figure \ref{fig:sketch}a) modulates the critical current. Variations of $I_c$ and the normal state conductance $G_N = 1/R_N$ with gate voltage are correlated, as seen previously by others \cite{Doh}. The junction is long compared to the superconducting coherence length in Nb, $\xi_{\mathrm{Nb}} \ll L \sim 200\; \mathrm{nm}$. We estimate an electronic mean free path $l_e$ on the order of $100\; \mathrm{nm}$, resulting in an intermediate regime between ballistic and diffusive transport. The mini-gap in the nanowire is determined \cite{thouless_energy_0,thouless_energy_1} by the Thouless energy $E_{\mathrm{Th}} = \hbar D/L^2$. Here, $D= l_e v_F / 3$ is the electron diffusion constant in the nanowire, and $v_F$ is the Fermi velocity. Using an estimated value for the Fermi energy $E_F \sim 150\; \mathrm{meV}$, we obtain $E_{\mathrm{Th}} \sim 0.5 \Delta_{\mathrm{Nb}}$. As discussed in the Supplementary Information, the superconducting coherence length in InAs, $\xi_{\mathrm{InAs}}$, is limited by dephasing to the inelastic scattering lengthscale, $l_{in}$. Using values for $l_{in}$ from magnetotransport studies on similar nanowires \cite{phase_coherence_length}, we estimate $\xi_{\mathrm{InAs}} \sim 250 - 500\; \mathrm{nm}$.\\
\indent The junction critical current $I_c$ was measured in perpendicular and axial magnetic fields, $B_\perp$ and $B_{\parallel}$ respectively. Below we discuss the behaviour of $I_c$ in each field direction.
\subsection*{Perpendicular magnetic field}
The perpendicular field $B_\perp$ was applied in the plane of the substrate and at an angle $2^\circ \pm 3^\circ$ with respect to $\bo {\hat{y}}$, the perpendicular direction to the nanowire axis. A complete penetration of the magnetic field into the normal section of the junction is assumed, and the screening of the field by Josephson supercurrents neglected. $I_c$ is found to have a monotonic, quasi-Gaussian decay with $B_\perp$ (Figure \ref{fig:sketch}b). The field was not increased beyond $B_\perp = 0.55\; \mathrm{T}$, where the switching transition becomes too weak for $I_c$ to be determined.\\
\indent A similar, monotonic behaviour of $I_c$ in a perpendicular field was observed previously for InAs and InN nanowire junctions \cite{proximityEffectInAs, proximityEffectInN}. In the narrow planar junction limit with perfect interfaces, $I_c$ is expected to exhibit a Gaussian decay \cite{narrow_junction_theory_01, narrow_junction_theory_02}, $I_c (B) = I_c (0) e^{-0.238 (\Phi / \Phi_0)^2}$, due to depairing. Here, $\Phi$ is the magnetic flux through the junction, $\Phi = B_\perp Ld$, $d$ the nanowire diameter, and $\Phi_0$ the (superconducting) flux quantum, $\Phi_0 = h/(2e)$. The experimental data, however, is best fit to a curve $I_c (B) = I_c (0) e^{-0.499 (\Phi / \Phi_0)^2}$. This discrepancy between the numerical factors can be explained in terms of non-ideal interfaces \cite{Hammer_non-ideal-interface}, i.e. a contact transparency less than $1$ -- see Supplementary Information.
\subsection*{Axial magnetic field}
An axial magnetic field is perpendicular to the Rashba spin-orbit direction, and the junction is predicted to enter the topological phase for a certain range of $B_\parallel$ values. Here, we study the dependence of $I_c$ on $B_\parallel$. The axial field was applied at an angle $\sim 8^\circ \pm 4^\circ$ with respect to the nanowire axis. The misalignment in $B_\parallel$ is shown in the Supplementary Information to have little effect on fitting the experimental data to the model, so for simplicity in what follows we assume $B_\parallel$ to be aligned. As with the perpendicular field, a complete penetration of $B_\parallel$ into the normal section of the junction is assumed, and the screening of the field by Josephson supercurrents neglected.\\
\indent Since the Josephson current is aligned with the field direction, naively one would expect no interference effects, and simply a slow decrease in $I_c$ with field due to depairing in the Nb leads. However, an oscillating behaviour in $I_c$ vs. $B_\parallel$ is observed, as shown in Figure \ref{fig:sketch}c. In terms of the flux through the axial cross-section of the nanowire $\Phi_{\mathrm{nw}} = B_\parallel \pi d^2/4$, the oscillations do not exhibit $\Phi_0$-periodicity. Moreover, the shape and periodicity of the oscillations can be modified by changing the gate voltages.\\
\indent What is the underlying physics of the observed oscillations? Recent observations of an oscillatory critical current in InSb junctions are consistent with 0-$\pi$ transitions of the junction phase in the presence of spin-orbit and Zeeman effects \cite{yokoyama_spin_orbit_zeeman}. However, we estimate for the InAs junction studied here, the magnetic field at which $I_c$ is predicted to have a minimum due to a 0-$\pi$ transition is $\sim 7$ T, an order of magnitude larger than the values observed $\sim 0.6\; \mathrm{T}$ (see Supplementary Information). Instead, we explain the effect as a form of Josephson interference that is analogous to the well-known effect for wide planar junctions in a perpendicular field that produces a Fraunhofer pattern, but here the azimuthal velocity component of carriers occupying subbands of finite orbital angular momentum yields a phase of magnetic origin, in addition to the zero-field superconducting phase difference across the junction. For simplicity we consider conduction on a shell, motivated by the tendency in InAs nanowires to form a surface accumulation layer of electrons due to the presence of surface states. Furthermore we use classical paths on the shell to calculate the additional phase for a given angular momentum state. While this model is semi-classical and based on simplifying assumptions, it illustrates that orbital subband effects can dominate the behaviour of $I_c$ in an axial field for multi-band nanowire $SNS$ junctions. Good agreement is found between the model and experimental results using a physically reasonable set of fitting parameters.
\subsection*{Junction critical current in a shell conduction model}
Intrinsic InAs nanowires typically have surface band bending $\sim 100 - 200\; \mathrm{meV}$ \cite{band_bending_meas} due to the pinning of the Fermi energy above the conduction band at the nanowire surface \cite{inasAcc0, inasAcc1, inasAcc2}, and a corresponding surface accumulation of carriers . We model the InAs nanowire junction as a cylindrical shell 2-dimensional electron gas (2DEG) contacted by Nb leads in the geometry of Figure \ref{fig:sketch}a. The 2DEG is assumed to be at a radius $d/2$ from the nanowire center. We consider relaxing this assumption and allowing more realistic transverse wavefunctions later. For numerical calculations, we choose the gauge $\mathbf{A} = (d/4)B_\parallel \hat {\boldsymbol{\theta}}$ for the vector potential ($ \hat {\bo \theta}$ is the azimuthal direction in the $yz$ plane).\\
\indent The motion of an electron in the 2DEG can be decomposed into an axial degree of freedom along $\hat {\boldsymbol x}$ and an azimuthal degree of freedom along $\hat {\boldsymbol \theta}$. The allowed quantum states are characterized by an angular momentum quantum number $l = 0,\pm 1,\pm 2,\; \mathrm{etc}$. In the ballistic limit, classical trajectories of electrons in subband $l$ going from source to drain are spiral paths on the cylindrical shell with a winding angle $\theta_l$ equal to the azimuthal arc length divided by the radius. So, $\theta_l = (v_\theta(l)) \times (L/v_x) \times (2 / d)$, where $v_x, v_\theta$ are the axial and azimuthal velocities, respectively, and $v_\theta$ is explicitly a function of the quantum number $l$. Below we also consider trajectories in the presence of back-scattering. The Andreev process involves an electron and a retroreflected hole, both located in the conduction band of the normal section of the junction \cite{eroms2008}. Therefore, the hole is considered to be a time and charge-reversed electron, with the same effective mass $m^*$, mean free path $l_e$ and radial position. The relative motion of the electron and hole can be neglected \cite{Belzig_quasiclassical}, and the charge transport described in terms of an `Andreev pair' \cite{two-electron-picture} with charge $-2e$, which follows the same trajectory as the electron. As long as the center of mass of the Andreev pair follows a spiral trajectory, the interference effect should appear, even if the holes do not have the same wavefunctions as electrons. Schematic examples of classical trajectories are shown in Figure \ref{fig:sketch}a for $l=0$ and $l=1$ states.\\
\indent An Andreev pair traversing the junction acquires a gauge-independent phase $\phi = (2e/ \hbar) \int \mathbf{A} \cdot \mathrm d \mathbf{l}$, due to the $\hat{\bo \theta}$ component of its momentum perpendicular to the field. This follows from the Ginzburg-Landau formula for the gauge-invariant phase \cite{tinkham}. The line integral, taken from one Nb electrode to the other, depends only on the azimuthal angle $\theta$ between the start and finish points, and not on the details of the path. For a spiral trajectory with winding angle $\theta_l$, one obtains $\phi = \frac {\Phi_{\mathrm {nw} }}{\Phi_0} \theta_l$. To our knowledge, the current-phase relationship (CPR) of a semiconducting nanowire $SNS$ junction has not been experimentally determined. Since we have $L \lesssim \xi_{\mathrm{InAs}}$, a sinusoidal CPR is assumed. The phase difference $\gamma$ between the Nb leads is assumed to be independent of the position along $\hat{\bo{ y}}$, since the junction width is similar in order to the Nb coherence length, $d \sim \xi_{Nb}$, corresponding to the narrow junction limit for a planar junction.\\
\indent Similar to the case of a wide planar junction in a perpendicular field \cite{Barone1} where the superconducting phase depends linearly on the position along the junction width, here the phase due to an axial field is linear in the winding angle $\theta$. Using the sinusoidal CPR, we define an angular supercurrent density $J(\theta) = J_c(\theta) \mathrm{sin} \left( \frac{\Phi_{\mathrm{nw}}}{\Phi_0} \theta + \gamma\right)$, where $J_c (\theta)$ is the critical current density. The supercurrent is obtained \cite{Barone1} by integrating the supercurrent density over $\theta$:
\begin{equation}
I ( \Phi_{\mathrm{nw}}) = \int _{-\infty}^{\infty} J(\theta) \mathrm d \theta,
\end{equation}
and the critical current is the maximum supercurrent over the junction phase $\gamma$:
\begin{equation}
I_c (\Phi_{\mathrm{nw}}) = \max\limits_{\gamma \in \left[ 0, 2\pi \right) } \left( I \right)=  \left| \int_{-\infty}^\infty J_c(\theta) \mathrm{e}^ {\left({ i \frac{ \Phi_{\mathrm{nw}} }{ \Phi_0 } \theta} \right)} \mathrm d\, \theta \right|.
\label{Eq:critial_current}
\end{equation}
Since $I_c$ and $J_c$ are related as a Fourier pair, $J_c (\theta)$ can be thought of as a spectral density that is a function of winding angle.
\subsection*{Spectral density of $I_c$ oscillations}
We model the spectral density $J_c(\theta)$ (see Eq. \ref{Eq:critial_current}) as a weighted sum of Gaussian functions in order to satisfy the following properties: (i) The subband $l$ contributes a peak to $J_c$ at its winding angle $\theta_l$, (ii) $J_c$ is proportional to the normal state conductance $G_N$, as required for an $SNS$ junction, (iii) $J_c$ goes to zero for large $|\theta|$, where the trajectory length is much longer than the phase coherence length $\xi_{\mathrm{InAs}}$. We write
\begin{equation}
J_c (\theta) = J_{\mathrm{max}} (\theta) \sum_l \frac {n_l}{\sigma \sqrt{2 \pi}} \mathrm{exp} \left( - \frac{ (\theta - \theta_l)^2} {2 \sigma ^2} \right),
\label{Eq:J_sum_of_peaks}
\end{equation}
where $n_l$ is the number of radial subbands occupied with angular momentum $l$. Here we allow $n_l >1$ in the spirit of later relaxing the assumption of strictly two-dimensional shell conduction so that higher radial excitations are possible. $J_\mathrm{max}$ takes into account the suppression of critical current density for large $|\theta|$, and is calculated from the Usadel equations in the Supplementary Information. The Gaussian peak width is determined by the parameter $\sigma$.\\
\begin{figure}[h!]
\centering
\includegraphics[width=0.48\textwidth]{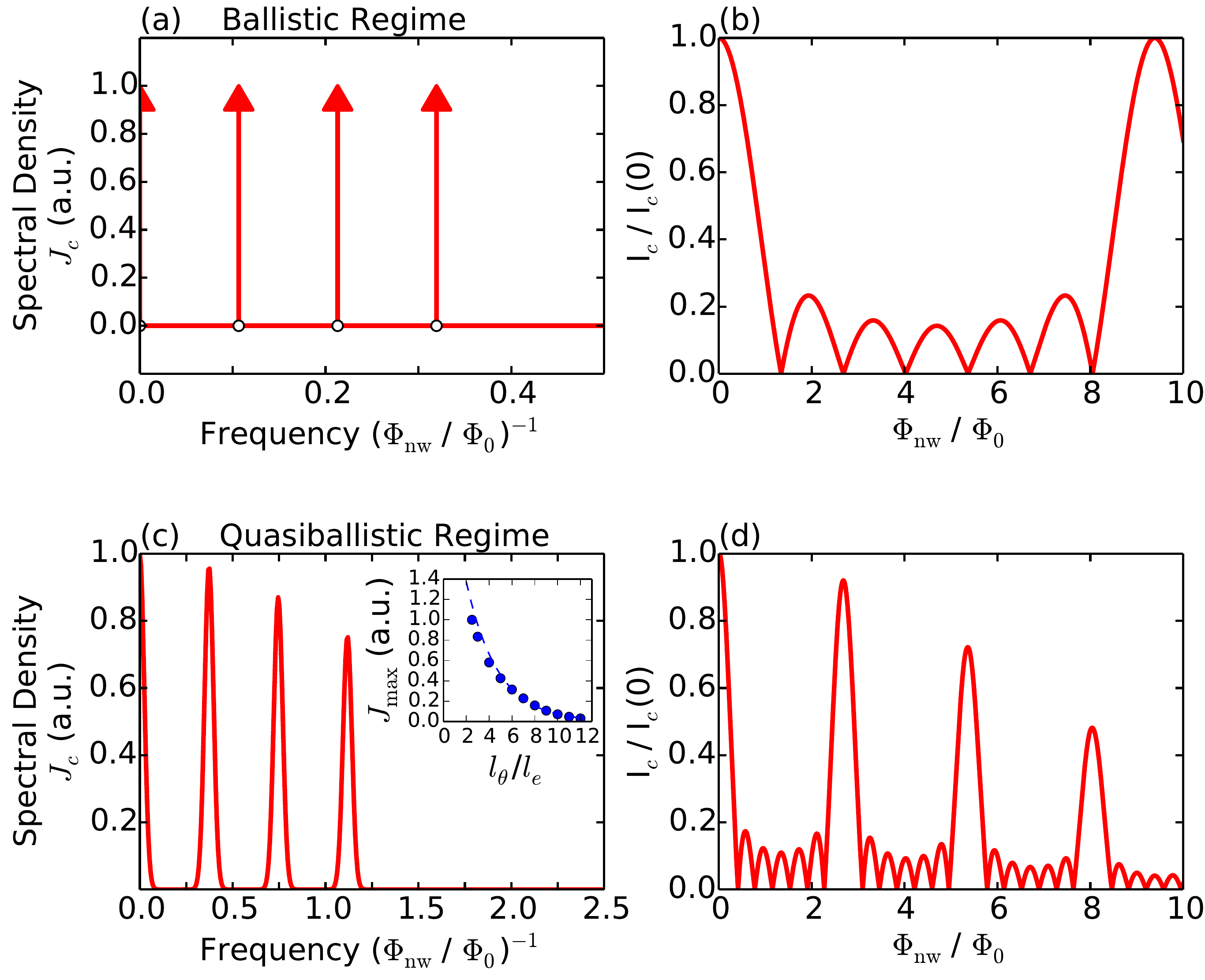}
\caption{\label{fig:model}
Spectral densities (left) and critical currents (right) for the examples given in the text, for (a,b) the ballistic regime, and (c,d) the quasiballistic regime. Although the spectral density is explicitly a function of the winding angle $\theta$, here it is plotted versus the magnetic frequency $\Phi_{\mathrm{0}}/\Phi_{\mathrm{nw}}$, since the inverse quantity $\Phi_{\mathrm{nw}}/\Phi_{\mathrm{0}}$ is the conjugate variable to $\theta$ in Eq. \ref{Eq:critial_current}. 
(a,b) A uniform $J_\mathrm{max}(\theta) = J_0$ is used in this example. The resulting $I_c$ is the absolute value of the sum of cosines in Eq. \ref{Eq:sum_of_cos}b.
(c,d) In the presence of back-scattering, the peaks in $J_c$ move to higher frequencies (Eq. \ref{Eq:qb}) as the carriers spend more time in the junction and accumulate more phase. Note that the frequency scales in (a) and (c) are not the same. The higher frequency peaks in $J_c$ are suppressed by $J_\mathrm{max} (\theta)$, which falls off with increasing winding angle, or equivalently, frequency. The oscillations in $I_c$ are qualitatively similar to the ballistic case, but have a shorter period, and attenuate with magnetic flux due to the broadening $\sigma$ of the $J_c$ peaks. (c, Inset) $J_{\mathrm {max}}$, as calculated from the Usadel equations, vs. $l_\theta / l_e$. Here, $l_\theta$ is the length of a spiral path with winding angle $\theta$, and an inelastic scattering length $l_{in} = 5l_e$ is chosen arbitrarily. The dashed line is an exponential fit to the $l_\theta > l_{in}$ region. Longer paths contribute less to $J_c$ because of dephasing, resulting in the suppression of the peak amplitudes with increasing magnetic frequency seen in panel (c). The following parameters (defined in the text) were used in these simulations: $d = 63\; \mathrm {nm}$, $L = 200\; \mathrm{nm}$, $l_e = 80\; \mathrm{nm}$, $l_{in} = 400\; \mathrm{nm}$, $E_F = 150\; \mathrm{meV}$, $n_l = 1$ for $|l| \leq 3$, and $n_l = 0$ otherwise.}
\end{figure}
\indent Let us consider first the ballistic limit $L \ll l_e$. One would expect sharp peaks (i.e.  $\sigma \rightarrow 0$) in $J_c (\theta)$ at $\theta = \theta_l$ for any subband $l$ that is occupied. In this limit, $v_x$ approximately equals the Fermi velocity of the electron, $v_x \simeq v_F = \sqrt{2E_F/m^*}$, and $v_\theta (l) = 2 \hbar l / (m^* d) \ll v_x$. We use $m^* = 0.023m_e$ as the effective mass in InAs. As an illustrative example, let $J_{\mathrm {max}}(\theta) = J_{0}$ for all $\theta$, and let $n_l$ = 1 for $|l| \leq 3$, and zero otherwise. Then, the critical current density is a sum of Dirac delta functions, and we have
\begin{subequations}
\begin{align}
J_c (\theta)& = J_{0} \sum_{l=-3}^3 \delta (\theta - \theta_l),\\
I_c (B_\parallel) & = J_{0} \left| \sum_ {l=-3}^{3} \mathrm{exp} \left( i \frac{\Phi_\mathrm{nw}}{\Phi_0} \theta_l \right) \right| \nonumber \\
\, & = J_{0} \left| 1 + 2 \sum_{l=1} ^ 3 \mathrm{cos} \left( \left( \frac{eLl}{m^* v_F} \right) B_\parallel \right) \right|.
\end{align}
\label{Eq:sum_of_cos}
\end{subequations}
Even though $\Phi_{\mathrm{nw}} \propto d^2 $, $I_c (B)$ does not depend on $d$ because for a fixed angular momentum, $v_\theta \propto 1/d $ so $\theta_l \propto 1/d^2 $ which cancels with the $d^2$ dependence of flux. This shows the interference effect is not sensitive to the radial position of carriers in the ballistic regime. In the limit $d\rightarrow 0$, states with finite angular momentum become very high in energy and will not be populated.\\
\indent In Figure \ref{fig:model}a is plotted $J_c (\theta)$ and $I_c(B_\parallel)$ for the example above (Eq. \ref{Eq:sum_of_cos}). Note that the $\theta_l$ which determine the peak positions in $J_c(\theta)$ depend on device-specific parameters $L, v_F$, and $d$, and can take on any value. Thus, the periodicity seen in $I_c$ versus $\Phi_{\mathrm{nw}}$ generally does not correspond to an integer multiple of $\Phi_0$.\\
\indent We extend the model for $J_c$ to the quasiballistic regime $l_e \lesssim L$ by invoking the following assumption: the electron undergoes back-scattering events along $\hat {\boldsymbol x}$ only. This is justified as long as the scattering does not substantially mix the orbital angular momentum states, such as when the scattering potential does not explicitly depend on the azimuthal position \cite{RacecPRB2009}. Therefore, in a scattering event $v_x \rightarrow -v_x$, but $v_\theta$ is unchanged, so on average the particle spends more time in the junction, accumulating more phase. Noting that in the ballistic limit we have $v_x \simeq v_F$, the drift velocity along the nanowire axis now becomes $v_x \simeq v_F \times l_e / (l_e + L)$. This follows from the scaling of the conductance $G$ from the Landauer-B{\"u}ttiker quantum conductance value by the factor $l_e / (l_e + L)$, as discussed in Refs. \cite{InAs_transistor, Datta_book}. Therefore, for quasiballistic transport due to back-scattering, the winding angle $\theta_l$ is taken as
\begin{equation}
\label{Eq:qb}
\theta_l = (\frac{L + l_e}{l_e})(\frac{L}{v_F}) (\frac{2}{d}) v_\theta (l).
\end{equation}
In Figure \ref{fig:model}b we plot $J_c (\theta)$ in the quasiballistic regime, corresponding to the example where $n_l$ = 1 for $|l| \leq 3$ and is zero otherwise. Using $l_e$ values typical for these nanowires \cite{InAs_transistor}, we obtain $(L + l_e)/l_e \sim 3.5$, so the peaks in $J_c(\theta)$ appear at higher frequencies (inverse magnetic flux) compared to the ballistic case. The effect of $J_\mathrm{max}(\theta)$, calculated from the Usadel equations, is taken into account. It suppresses the higher frequency peaks of $J_c$, since these correspond to longer classical paths of the carriers that will experience greater dephasing. Intuitively, one expects a broader distribution of $J_c(\theta)$ about each $\theta_l$-centered peak when back-scattering occurs, and this broadening is parametrized by $\sigma$. A finite $\sigma$ suppresses the recurrences in $I_c$; as $\sigma$ is increased, the maxima in $I_c$ drop off with increasing magnetic flux. The experimental data is fit below using this quasiballistic model. It is shown in the Supplementary Information that the interference effect is not sensitive to the radial position of the carriers, $d/2$, up to a rescaling of the broadening parameter $\sigma$, and of the envelope function $J_{max}$.
\subsection*{Fitting to the data}
\begin{figure}[t!]
\centering
\includegraphics[width=0.48\textwidth]{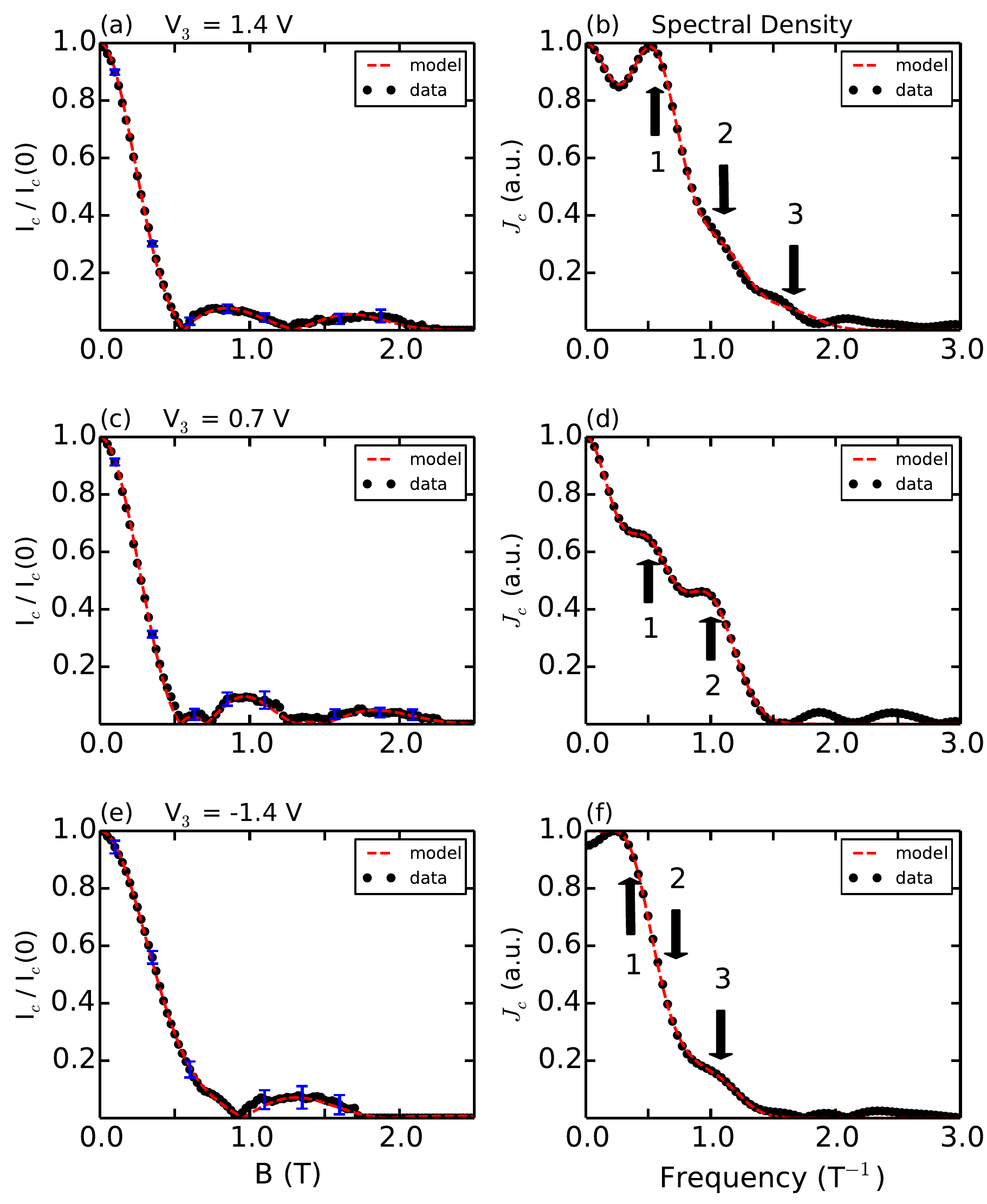}
\caption{\label{fig:fits} Theoretical fits to the experimental data for (a, b) $V_3 = 1.4\; \mathrm{V}$, (c, d) $V_3 = 0.7\; \mathrm{V}$ and (e, f) $V_3 = -1.4\; \mathrm{V}$. (a, c, e) The experimental critical current (solid circles) and model fits (dashed lines) are overlaid. The data is truncated above magnetic fields for which the switching transition becomes too weak for $I_c$ to be determined. (b, d, f) Spectral densities of the experimental data determined by Fourier transformation of $I_c$ (solid circles), and the best-fit spectral densities from the model (dashed lines) are shown. To produce the model curves, Eq. \ref{Eq:J_sum_of_peaks} is fitted to the low-frequency region $ < 2\; \mathrm{T}^{-1}$ with the parameters given in Table \ref{TAB:fitpar}. Arrows indicate the frequencies of the peaks corresponding to each subband of orbital angular momentum $\hbar |l|$, for $|l|=1,2,3$ ($l=0$ gives a peak at zero frequency). For clarity, we only show every fourth experimental data point, and representative error bars on selected data points.}
\end{figure}

In Eq. \ref{Eq:J_sum_of_peaks} we have modelled $J_c (\theta)$, the spectral density of $I_c$. In order to fit the model to the experimental data, spectral densities are calculated numerically from the $I_c (B)$ data shown in Figure \ref{fig:fits}, for three values of gate voltage $V_3$. A fast-Fourier transform is taken of a signal identical to $I_c(B)$, but that changes sign at each node so that each becomes a zero crossing. Eq. \ref{Eq:J_sum_of_peaks} is then fitted to this Fourier transform, with results shown in Figure \ref{fig:fits}, and discussed below.
\begin{table}
\begin{tabular}{c | c c c c c c }

$\;$ V$_3$ $\;$&$\;$ $l_e$ $\;$&$\;$ $\sigma$ $\;$&$\;$ $n_0$ $\;$ &$\;$ $n_{\pm 1}$ $\;$& $\;$ $n_{\pm 2}$ $\;$&$\;$ $n_{\pm 3}$ $\;$\\

(V) & (nm)\\  \hline \hline
 1.4 & 83.1  & 0.9 & 2.0 & 2.1 & 0.7 & 0.2 \\
 0.7 & 94.9  & 0.9 & 2.0 & 1.3 & 1.0 & 0.0 \\
-1.4 & 168.0 & 0.8 & 1.0 & 1.1 & 0.3 & 0.2 \\ 

\end{tabular}
\caption{\label{TAB:fitpar} The parameters used in the fits of Figure \ref{fig:fits}; $l_e$ is the mean free path, $n_l$ are the subband occupation numbers, and $\sigma$ describes the broadening of the angular spectral density function. Since the model only constrains the ratios $n_l / n_{l+1}$, the $n_l$ here are adjusted to be consistent with the experimental normal state conductance, $G_N$, at each gate voltage. The following parameters (defined in the text) are used but not varied between the three data sets: $d = 63\; \mathrm{nm}$, $L = 200\; \mathrm{nm}$, $E_F = 150\; \mathrm{meV}$, $l_{in} = 400\; \mathrm{nm}$. }
\end{table}
\section*{Discussion}
Here we discuss the best-fit parameters we obtained by fitting the model to the experimental data (Table \ref{TAB:fitpar}). The most important variation between the three data sets is the occupation number of angular momentum subbands, $n_l$, which is modulated by the applied gate voltage. As $V_3$ is made more negative, the fits are consistent with fewer subbands being occupied. The subbands with higher $l$ values depopulate first, in a manner roughly consistent with the expected shell-filling structure of the nanowire \cite{schapers_shell_filling_10, schapers_shell_filling_13, greg-mc}. For a given data set, only the ratios $n_l / n_{l+1}$  matter when calculating $I_c (B) / I_c(0)$. We have adjusted the $n_l$ so that the experimental normal state conductance $G_N$ at each gate voltage is consistent with the estimated number of occupied subbands. Another notable feature in Table \ref{TAB:fitpar} is an apparent increase in elastic mean free path $l_e$ as the gate voltages are made more negative. In the real nanowire junction, the radial wavefunction is not confined to a shell but rather is distributed through the nanowire cross section, with some bias towards the surface due to surface band bending. Higher $l$ states have a radial expectation value closer to the nanowire surface and are expected to scatter more frequently due to surface defects, which could explain a shorter mean free path when more channels contribute to transport. Note that the fitting involves several free parameters, and the set of best-fit parameters is not generally unique. The model is based on simplifying assumptions and considers shell conduction only, and also neglects spin-orbit and Zeeman effects. We therefore do not expect this model to capture the complete physics of the junction; however, it appears to explain the dominant mechanism of the critical current oscillations in the InAs device studied here.\\
\indent In conclusion, a nanowire $SNS$ junction was investigated that showed an unexpected modulation of critical current in an axial magnetic field. This result is understood by considering a novel type of Josephson interference due to orbital states in the multi-band InAs nanowire. Although we restricted the model to a cylindrical shell, the interference effect does not depend on the diameter $d$, up to a rescaling of the spectral density broadening parameter $\sigma$. The effect is therefore expected to be present for more general radial wavefunctions, including when electrons and holes forming Andreev pairs  have different radial wavefunctions. For a more realistic model, it will be necessary to calculate the Josephson interference based on quasiparticle wavefunctions rather than classical trajectories. Despite the simplicity of the semi-classical model considered here, it is able to reproduce the main features of the experimental data. Therefore, the interplay of orbital and magnetic effects should be carefully considered for semiconductor nanowire Josephson junctions, especially when searching for signatures of topological states.
\section*{Methods}
The bottom gate pattern was defined by electron beam lithography (EBL) on an undoped Si substrate with a 300 nm thermal oxide layer. 7 nm/14 nm layers of Ti/Au were deposited by electron beam evaporation to realize the gates, followed by atomic layer deposition of a 7 nm layer of dielectric Al$_2$O$_3$, and plasma-enhanced chemical vapour deposition of a 13 nm layer of dielectric SiN$_\mathrm{x}$. Next, molecular beam epitaxially grown InAs nanowires were mechanically deposited on the substrate. Details of the nanowire growth can be found in Ref. \cite{wire-growth}. Using scanning electron microscopy, we selected a nanowire positioned on the predefined gate pattern. The pattern for the superconducting contacts was defined by EBL. A 50 nm layer of Nb was deposited by dc sputtering at room temperature, preceded by Ar ion milling to achieve transparent InAs/Nb interfaces.\\
\indent The sample was wirebonded to a chip carrier and thermally anchored to the mixing chamber of a dilution refrigerator with a base lattice temperature of 25 mK. The junction was connected in a four-probe (current-voltage) setup. A dc currant bias was applied using a homemade voltage source, by dropping the output voltage across two sets of resistors, anchored at room temperature, and the mixing chamber temperature of the dilution refrigerator, respectively. The voltage response of the junction was measured using a voltage preamplifier.
\section*{Acknowledgements}
We gratefully acknowledge discussions with A. J. Leggett, B. Spivak, A. Burkov, R. Aguado, and E. Prada, and the technical support provided by the Quantum NanoFab facility. R. Romero provided technical assistance. We are grateful to B. Plourde and M. Ware (Syracuse University) for assistance with Nb deposition. This work was supported by NSERC, Canada Foundation for Innovation, the Ontario Ministry of Research \& Innovation and Industry Canada. G.W.H. acknowledges a WIN fellowship.
\section*{Author Contributions}
K.G. and J.B. designed the experiment. C.M.H. and R.R.L. performed nanowire growth. G.W.H., M.M. and K.G. fabricated the junction. J.B. and K.G. acquired the experimental data. K.G. performed data analysis and fitting. K.G. and M.H.A. developed the semiclassical model in consultation with J.B. K.G. and J.B. wrote the manuscript. The project was supervised by J.B.
\section*{Competing Financial Interests}
The authors declare no competing financial interests.

%\putbib[nanowire-interference]
%\end{bibunit}
%\bibliographystyle{h-physrev}
%\bibliography{nanowire-interference.bib}
%end main text
%%%%%%%%%%%%%%%%%%%%%%%%%%%%%%%%%%%%%%%%%%%%%%%%%%%%%%%%%%%%%%%%%%%%%%%%%%%%%%
%%%%%%%%%%%%%%%%%%%%%%%%%%%%%%%%%%%%%%%%%%%%%%%%%%%%%%%%%%%%%%%%%%%%%%%%%%%%%%
%Start Supplementary Information

%%%%%%%%%% Prefix a "S" to all equations, figures + reset counters %%%%%%%%%%
\makeatletter
\setcounter{equation}{0}  % reset counter
\newcounter{newfigure}		%use a new counter for supplementary figures -- needs to be manually incremented.
\renewcommand{\theequation}{S\arabic{equation}}
\renewcommand{\thefigure}{S\arabic{newfigure}}
\renewcommand{\bibnumfmt}[1]{[S#1]}
\renewcommand{\citenumfont}[1]{S#1}
\makeatother
%%%%%%%%%% Prefix a "S" to all equations, figures + reset counters %%%%%%%%%%
\begin{widetext} \label{SUPP-INFO}
\newpage
%\begin{bibunit}[h-physrev]
\begin{center}
\large \textbf{Supplementary Information: Josephson Interference due to Orbital States in a Nanowire Proximity Effect Junction}
\end{center}
%\tableofcontents
\section{Junction Characteristics}
\subsection{Tunability of the Supercurrent with Local Gate Voltages}
\label{subsec:tunability}
Figure \ref{fig:a-tune} shows the differential resistance $dV/dI$ of the junction vs. the bias current $I$ and the local gate voltage $V_3$ (see Figure 1 in the main text.) A clear zero-resistance regime can be seen at low bias. A voltage drop of magnitude $\lesssim 5\; \mathrm{\mu eV}$ develops across the junction as $I$ approaches the critical current $I_c$. This voltage has an exponential dependence on the bias current $I$, and is consistent with the occurrence of a combination of thermally activated phase slip events and quantum phase slip events, similar to the effect observed in Ref. \cite{supp_Sahu2009}.
\indent The voltages on the local gates can be used to tune the critical current of the junction. Of these, $V_3$ is the most effective. The variations of $I_c$ with $V_3$ are reproducible. A general trend is observed that $I_c$ is larger for more positive $V_3$. In Section \ref{subsec:Ic_Gn} we show that these variations are correlated with the variations of the normal state conductance, $G_N$, with $V_3$. We conclude that the normal section of the $SNS$ junction is indeed semiconducting, and the gate voltages tune the local potential in the nanowire, simultaneously tuning $I_c$ and $G_N$.
\begin{figure}[h!]
\stepcounter{newfigure}
\centering
\includegraphics[width=0.6\textwidth]{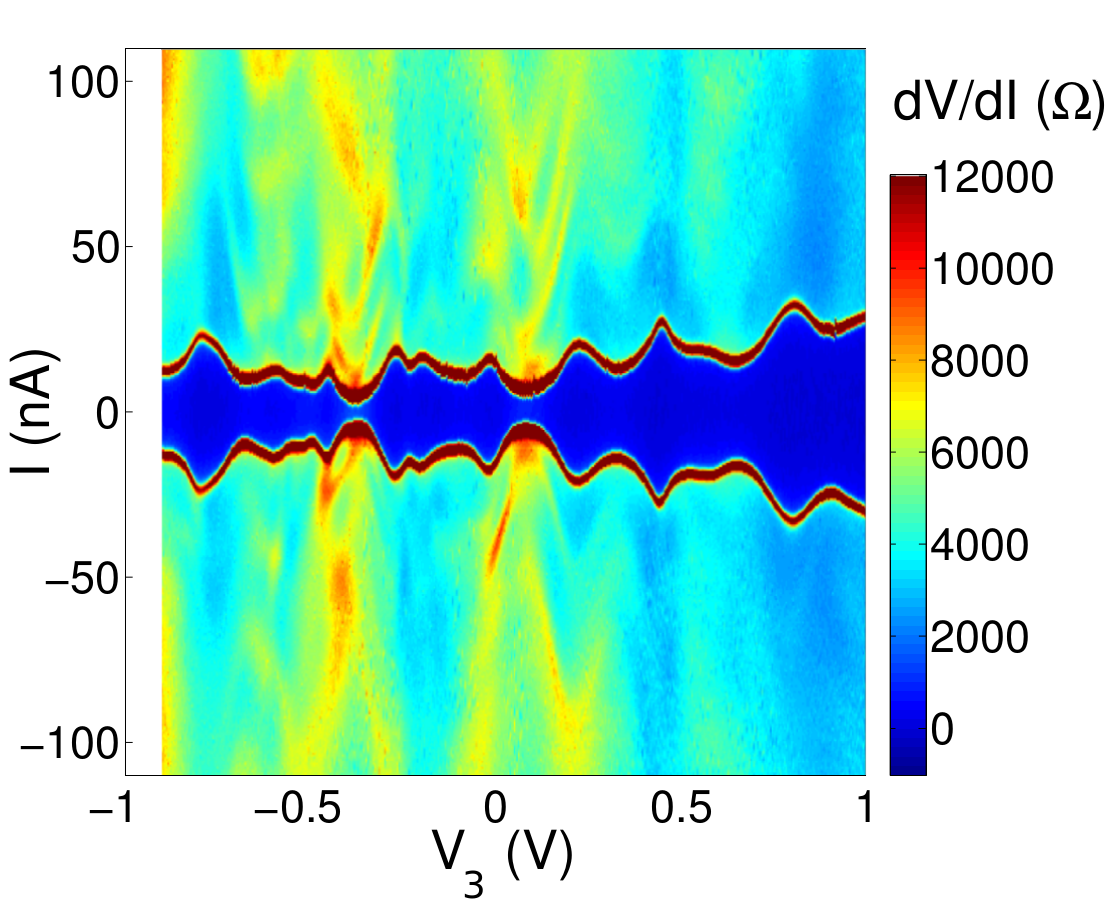}
\caption{\label{fig:a-tune} 
Differential Resistance $dV/dI$ in Ohms, vs. bias current $I$ and local gate voltage $V_3$. Here, the voltages on the two gates $V_3, V_4$ are being swept together from negative to positive values. The critical current shows reproducible variations with $V_3$.}
\end{figure}
\subsection{Contact Resistances}
\label{subsec:contactres}
In order to correctly identify the resistance of the nanowire section of the junction, the resistances of the Nb-InAs contacts need to be subtracted from the total measured resistance. Since the nanowire is contacted by two Nb leads, a four-point measurement of the contact resistances was not possible. Using the results in \cite{supp_barriers}, the barrier resistance $R_b$ was estimated to be no greater than $1\; \mathrm{k\Omega}$. In the main text, and in what follows, we consistently use the value $R_b = 800\; \Omega$ for each contact. The normal state resistance of the junction $R_N$ is related to the measured high bias current resistance value $R$ by the relation $R = (1 + 2r)R_N$, where $r = R_b / R_N$.  Typically for this device, the value for the ratio $r$ is $r \sim 0.15$. 
\subsection{Correlation of the Critical Current and Normal State Conductance}
\label{subsec:Ic_Gn}
The resistance $R$ of the junction was measured at a high bias voltage $V = 5\; \mathrm{mV} > 2\Delta_{Nb}/e$, where $\Delta_{Nb} = 1.2\; \mathrm{meV}$ is the superconducting gap in the Nb leads. The normal state resistance $R_N$ of the junction was obtained using the relation $R = R_N + 2R_b$, where $R_b = 800 \Omega$ is the resistance of the Nb-InAs barriers (see Section \ref{subsec:contactres}).\\
\indent In figure \ref{fig:c-Corr} we show the normal state conductance $G_N = 1/R_N$ and the critical current $I_c$ of the junction vs. the local gate voltage $V_3$. The variations of $G_N$ and $I_c$ with $V_3$ are correlated, suggesting a common physical origin. This has been previously observed for semiconductor nanowire $SNS$ junctions \cite{supp_Doh,supp_jesInAs,supp_proximityEffectInN,supp_proximityEffectInAs,supp_abay_icgn_ballistic}. For a nanowire junction in the ballistic regime, $I_c$ and $G_N$ are directly proportional \cite{supp_abay_icgn_ballistic}. This is not the case for the junction studied here, and can be attributed to non-ideal Nb-InAs interfaces (i.e. a contact transparency less than 1), which suppress the mini-gap and therefore the critical current of the junction \cite{supp_Hammer_non-ideal-interface}. The relative fluctuations in $I_c$ are larger than those in $G_N$, as seen in Doh, et al \cite{supp_Doh}. This can be attributed to an interplay between the phase coherent Andreev processes and conductance fluctuations resulting from potential fluctuations in the nanowire. 
\begin{figure}[h!]
\stepcounter{newfigure}
\centering
\includegraphics[width=0.6\textwidth]{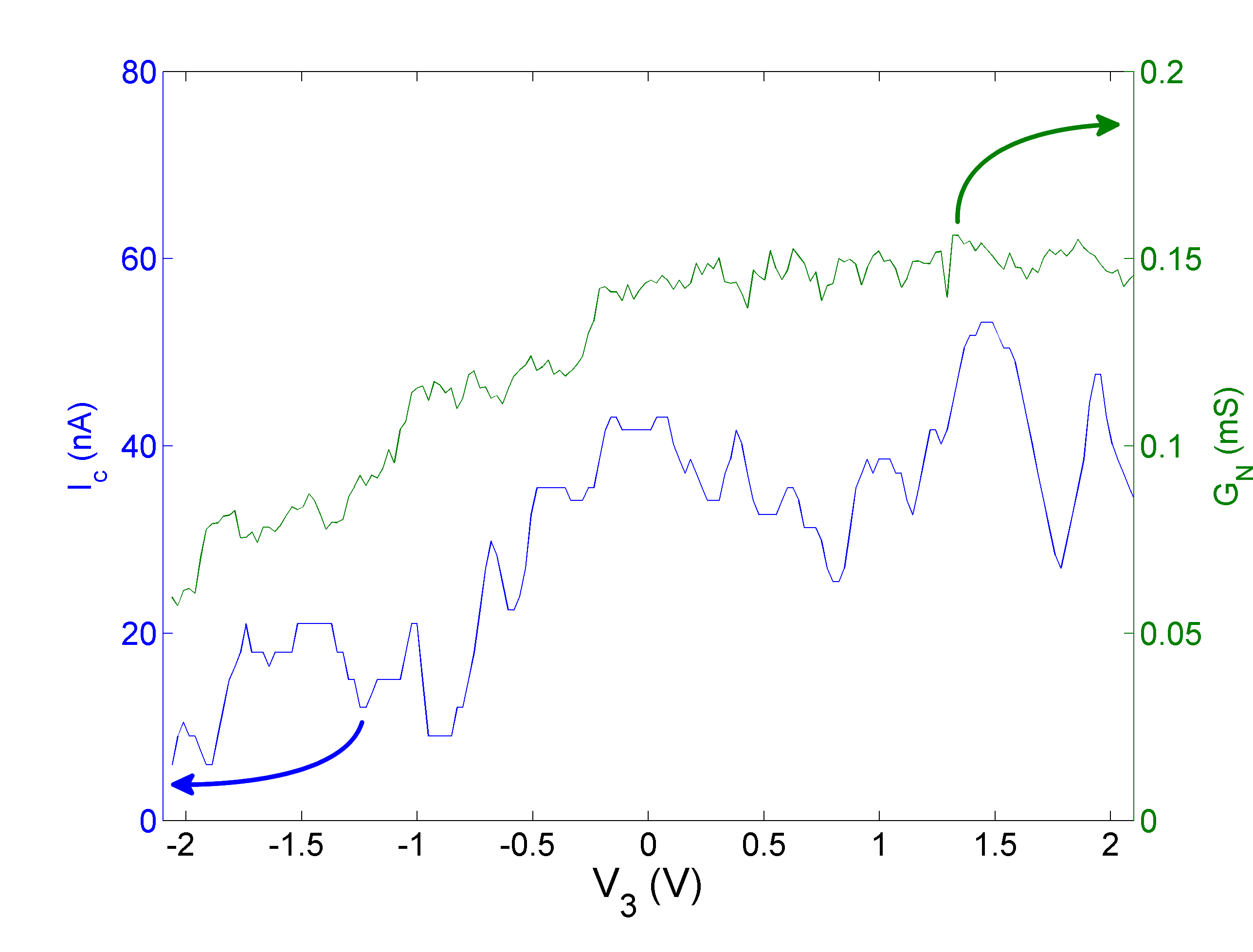}
\caption{\label{fig:c-Corr} 
Junction critical current $I_c$ (blue) and high bias current normal state conductance $G_N$ (green) vs the local gate voltage $V_3$. The variations of $I_c$ and $G_N$ with $V_3$ are clearly correlated, although a direct proportionality relation does not hold, due to non-ideal Nb-InAs interfaces.}
\end{figure}
\subsection{Multiple Andreev Reflections}
\label{subsec:MAR}
Signatures of Multiple Andreev Reflections (MAR) \cite{supp_andreev} are visible as peaks in the differential resistance $dV/dI$ of the junction, at bias voltages $V_n = 2\Delta_{Nb}/en$ for integer $n$. Here, $\Delta_{Nb} = 1.2\; \mathrm{meV}$ is the superconducting energy gap in the Nb leads.  The size of the peak at each $V_n$ varies with the local gate voltages, but the peaks are visible whenever there is a supercurrent present at zero bias voltage. The presence of this subharmonic gap structure \cite{supp_otbk_corrected} indicates phase coherence across the junction.\\
\indent In Figure \ref{fig:d-MAR} the junction differential resistance vs. bias voltage is shown for $V_3 = +2.1\;\mathrm{V}$. A supercurrent branch is present at zero bias voltage, and signatures of MAR are clearly visible at higher bias. As the bias voltage is increased, the Andreev process at the Nb-InAs interfaces is suppressed, increasing the junction resistance.
\begin{figure}[h!]
\stepcounter{newfigure}
\centering
\includegraphics[width=0.8\textwidth]{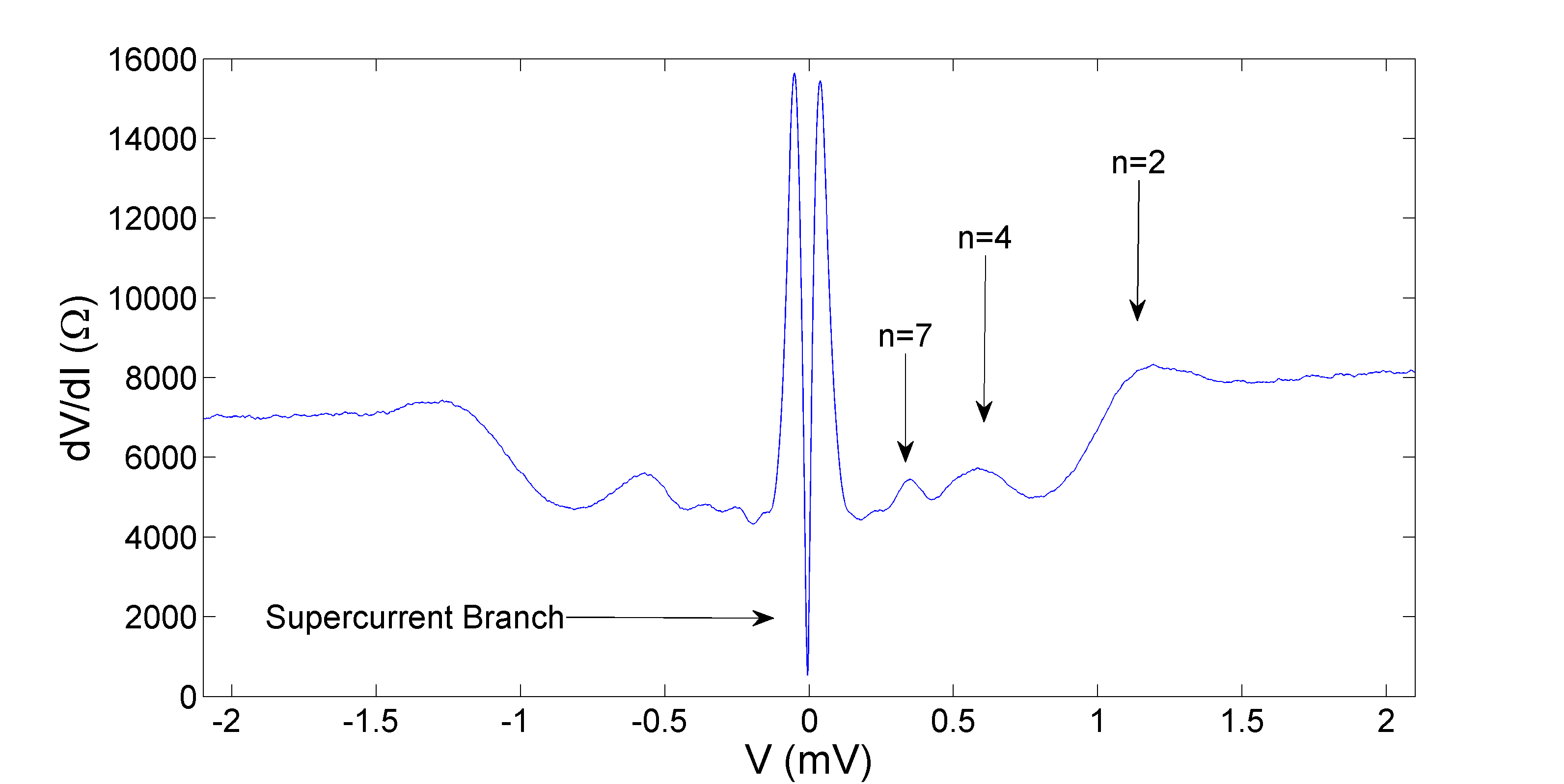}
\caption{\label{fig:d-MAR}
Junction differential resistance $dV/dI$ vs bias voltage $V$. A supercurrent branch can be seen at zero bias. Signatures of Multiple Andreev Reflections (MAR) are seen higher bias as peak in $dV/dI$. This indicates phase coherence across the junction. The arrows indicate the position and the number of MAR peak. For this value of the gate voltage, $V_3 = +2.1\; \mathrm{V}$, the bias voltage was not raised to $V > 2\Delta_{Nb} / e$ because of the low resistance of the junction. For other values of $V_3$, the $n=1$ MAR signature can be observed.}
\end{figure}
The current-voltage ($I$-$V$) trace corresponding to Figure \ref{fig:d-MAR} can be used to extract the excess current \cite{supp_otbk_corrected} due to the Andreev process, by making a linear fit to the high bias voltage regime. This yields $I_{exc} = 40.2\; \mathrm{nA}$ (Figure \ref{fig:d-excess}). This value, and the value for the high-bias conductance of the junction, are used to calculate the figure of merit product $eI_{exc} R_N /\Delta_{Nb} \simeq 0.5$. This value is inserted into the OTBK model \cite{supp_otbk_corrected} for the subharmonic gap structure, in order to get the scattering parameter $Z$ at the Nb-InAs interfaces. For $Z$ we get a value $\sim 0.75$, indicating a Nb-InAs contact transparency of $t \sim 0.65$.\\
\indent Notice in Figures \ref{fig:d-MAR}, \ref{fig:d-excess} the bias voltage was not be increased to $V > 2\Delta_{Nb} / e$, because the low resistance of the junction would result in a large current, and therefore a high Joule heating. However, for $V > 2\Delta_{Nb} / e$ we would expect the excess current to be even higher than the measured value. Therefore, the reported value $t \sim 0.65$ is a lower bound of the actual contact transparency of the junction. For other values of the gate voltage $V_3$, the MAR signature at $V = 2\Delta_{Nb} / e$ can be observed.
\begin{figure}[h!]
\stepcounter{newfigure}
\centering
\includegraphics[width=0.6\textwidth]{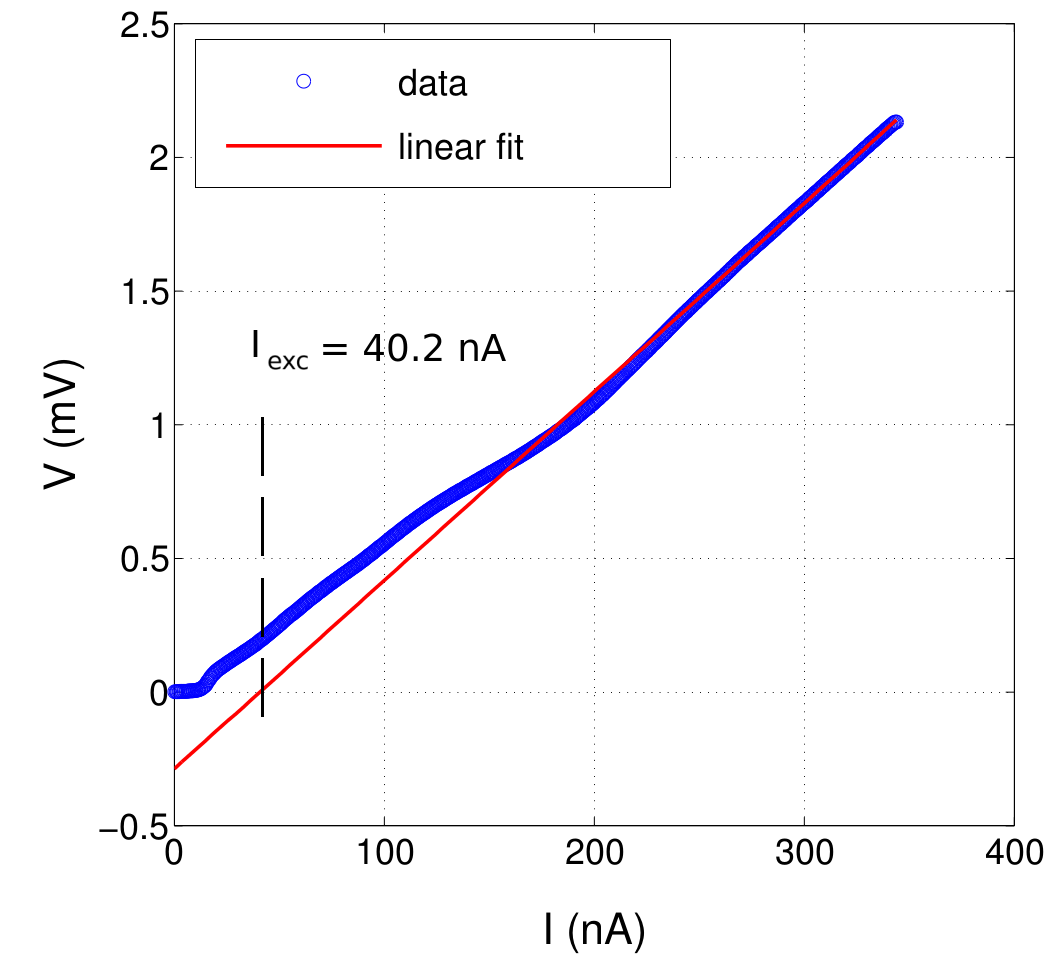}
\caption{\label{fig:d-excess}
Current-Voltage trace of the junction (blue circles), at the gate voltage $V_3 = +2.1\; \mathrm{V}$. A linear fit to the high bias voltage regime (solid red line) is used to extract the excess current $I_{exc} = 40.2\; \mathrm{nA}$. This value indicates a Nb-InAs contact transparency $t \sim 0.65$.}
\end{figure}
\subsection{Superconducting Coherence Length}
\label{subsec:coherence-length}
Thermal fluctuations and inelastic scattering result in the dephasing of Andreev pairs, limiting the superconducting coherence length in the normal section of an $SNS$ junction \cite{supp_dubos2001, supp_Hammer_non-ideal-interface}. For the junction studied here, the superconducting coherence length in the InAs nanowire is $\xi_{\mathrm{InAs}} = \mathrm{min}(l_T, l_{in})$, where $l_T = \sqrt{\hbar D / (2 \pi k_B T)}$ is a characteristic thermal length, $D$ the diffusion coefficient, $k_B$ the Boltzmann constant, $T$ the (electron) temperature, and $l_{in}$ the inelastic scattering length. Using an estimated $T \sim 100 \mathrm{mK}$, we calculate $l_T$ on the order of a micrometer.\\
\indent The inelastic scattering length in InAs nanowires, similar to the one studied here, was measured in Ref. \cite{supp_blomers2011} to be $l_{in} \sim 250 - 500\; \mathrm{nm}$. This indicates that $\xi_{\mathrm{InAs}}$ is limited by $l_{in}$. In our calculations we consistently use $\xi_{InAs} = 400\; \mathrm{nm}$, longer than the junction length $L \simeq 200\; \mathrm{nm}$.
\section{Junction Critical Current in a Perpendicular Magnetic Field}
\label{sec:perp}
\indent The Junction critical current exhibits a monotonic, quasi-Gaussian decay as the perpendicular magnetic field, $B_\perp$, increases (see Figure 1b in the main text). Here we concentrate on the theoretical description of this behaviour.\\
\indent In Ref. \cite{supp_Hammer_non-ideal-interface}, the behaviour of the critical current of a long, diffusive planar $SNS$ junction in a perpendicular field was studied using the quasiclassical Green's function method. The roles of non-ideal N-S interfaces were considered. The perpendicular field results in a pair-breaking mechanism equivalent to spin-flip scattering, which is responsible for the decay of the critical current $I_c$ with $B_\perp$. For a narrow junction (with width $w$ smaller than the superconducting coherence length $\xi$) with perfect interfaces, the following regimes were found: (i) When the junction is very long, $w \ll L$, the Usadel equation for the Green's function can be solved analytically, resulting in a Gaussian decay as a function of the magnetic flux $\Phi$ through the junction: $I_c(\Phi) = I_c(0) \mathrm{exp}(-0.238(\Phi / \Phi_0)^2)$. Here, $\Phi_0$ is the (superconducting) flux quantum. (ii) If the width is lesser but of the order of the length, numerical methods are required; however, $I_c$ still decays, and the decay is approximately Gaussian. Aside from the interplay of the orbital states and the perpendicular field (see Section \ref{sec:interplay}), the nanowire junction studied here is analogous to a planar junction with $w$ given by the diameter $d = 63\; \mathrm{nm}$. The junction is narrow, $d \lesssim \xi_{InAs} \sim 250 - 500\; \mathrm{nm}$. The junction length is given by $L = 200\; \mathrm{nm}$. Thus, the model for a planar junction with $w \lesssim L$ applies. Calculations reported in Ref. \cite{supp_geometry-mag} indicate that in this limit, a quasi-Gaussian decay of $I_c$ with $B_\perp$ is expected, with a tail at high $B_\perp$ slightly above the Gaussian curve. This may explain the why the experimental critical current in figure 1b develops a tail when $B_\perp > 0.3\; \mathrm{T}$.\\
\indent The experimental $I_c$ is fit to a Gaussian $I_c(\Phi) = I_c(0) \mathrm{exp}(-0.499(\Phi / \Phi_0)^2)$, which decays faster by a factor of $2.1$ than theoretically predicted in case (i) above. This can be explained by taking into account non-ideal Nb-InAs interfaces. As discussed in \cite{supp_Hammer_non-ideal-interface}, the behaviour of the junction is critically dependent on contact resistances. The main effect of non-ideal interfaces is to increase the dwell time of a the Andreev pair in the semiconducting section of the junction. The cumulative spin-flip depairing effect causes $I_c$ to decay faster. We use an estimated $R_b = 800\; \mathrm{\Omega}$ for each contact, as discussed in Section \ref{subsec:contactres} . This implies a resistance ratio of $r = R_b / R_N = 0.15$, which can account for rescaling the decay by a factor of $1.6$, see \cite{supp_Hammer_non-ideal-interface}. An additional contribution can come from an asymmetry between the contacts, which would further increase the decay rate.\\
\section{Interplay of Orbital States and the Perpendicualr Magnetic Field}
\label{sec:interplay}
A natural question to ask is if the interference effect due to the interplay of orbital states and the external field appears when the magnetic field is applied in a perpendicular rather than an axial direction.\\
\indent Let the nanowire axis be along the $\hat{\bo x}$ direction (figure 1a in the main text). Let a magnetic field $\bo B = -B_\perp \hat{\bo y}$ be applied. For the vector potential we choose $\bo A = (B_\perp z / 2) \hat {\bo x}$. The main effect of this field is to attenuate the critical current with $B_\perp$ through a depairing mechanism (Section \ref{sec:perp}). We ask here if the phase picked up by spiral paths on the circumference of the nanowire can result in an oscillatory behaviour on top of this attenuation.\\
\indent Consider a spiral path with winding angle $\theta$. Let us assume no back-scattering for the moment (ballistic regime). The phase picked up by this path is $\phi = 2e/ \hbar \int \bo A \cdot \mathrm{d} \bo l$, where the line integral is taken along the path. Since the path is a simple spiral, the height $z$ can be determined as a function of the position $x$  along the junction length as $z = r\; \mathrm{cos}\: (x \theta /L)$. Here, $r = d/2$ is the radius of the nanowire and $L$ the length of the junction. So the phase is
\begin{equation}
\phi = (e B_\perp r/\hbar) \int_0^L \mathrm{cos}\: (x\theta / L) \mathrm{d}x = 2\pi \left( \Phi / \Phi_0 \right) \left( \frac{\mathrm{sin}\: (\theta)}{\theta} \right),
\label{Eq:phase_perp_interplay}
\end{equation}
where $\Phi = B_\perp d L $ is the flux through the junction and $\Phi_0$ is the (superconducting) flux quantum. Eq. \ref{Eq:phase_perp_interplay} indicates a qualitative difference versus the case of an axial magnetic field:  as the perpendicular component of the momentum changes direction with respect to the applied field, the phase picked up by the spiral path changes sign, so the accumulated phase cancels with itself. Specifically, the phase accumulated on parts of the spiral path with $z > 0$ cancels out with that of parts with $z < 0$. Whereas the phase picked up by long spiral paths is suppressed, for a junction in the ballistic regime, the winding angles of the angular momentum subbands with $l = -3,-2,\ldots,3$ are in the range $(- \pi,\pi)$, i.e. on the main lobe of the sinc function in Eq. \ref{Eq:phase_perp_interplay} ($l$ is the angular momentum quantum number). This means that angular momentum subbands can pick up an appreciable phase. Therefore, the occurrence  of an interference effect may be possible for a short, ballistic junction.\\
\indent Now let us consider the quasiballistic regime by introducing back-scattering. Notice that $\bo A \cdot \mathrm{d} \bo l$ changes sign upon each back-scattering event. This has a randomizing effect on the total phase, as the phase now depends on the details of the scattering events. Since the phase $\phi$ picked up between two scattering events is not small ($|\phi| \sim \pi/2$), after several scattering events we expect the phase to be completely randomized, and no interference effect to survive. Similar to previously studied cases of a randomly distributed phase due to diffusive paths \cite{supp_path-inteference-arxiv,supp_geometry-mag}, the effect of this random phase is estimated to be a suppression of the total supercurrent, contributing the suppression of $I_c$ with $B_\perp$ on top of the pair-breaking mechanism discussed in Section \ref{sec:perp}.\\
\indent In summary, for the quasiballistic junction studied experimentally we do not expect an interference effect due to the interplay of orbital states and a perpendicular field, because of the randomizing effect of back-scattering. However, in a ballistic junction short enough that the self-cancelling of the phase does not dominate, an interference effect may occur.
\section{Spin-Orbit and Zeeman Effects in the Junction}
\label{sec:0-pi}
Yokoyama et al. \cite{supp_so_zeeman}  theoretically studied the interplay of spin-orbit and Zeeman effects in an InSb nanowire $SNS$ junction. They found that, in the presence of a magnetic field $B$, the different spin components of Andreev pairs pick up different amounts of superconducting phase, resulting in the splitting of the Andreev Bound States in the junction. Consequently, as $B$ is increased from zero, the junction critical current $I_c$ is reduced. At a certain value for $B$, the junction abruptly undergoes a so-called 0-$\pi$ transition, wherein the phase $\phi$ across the junction in its ground state shifts from $\phi = 0$ to $\phi = \pi$. This results in a cusp (minimum) in the critical current, and, upon further increasing the field $B$, $I_c$ increases until it recovers value close to its original value at $B = 0$. The cycle repeats upon further increasing $B$, resulting in an oscillatory $I_c$. This effect can dominate the behaviour of $I_c$ for InSb nanowire junctions in the presence of a magnetic field. We ask if a similar physical picture can be responsible for the observed oscillations of $I_c$ vs $B_\parallel$, in the InAs nanowire junction studied here.\\
\indent In terms of a magnetic parameter $\theta_B = |g \mu_B B| L / ( \hbar v_F)$, the cusps of the critical current are predicted to happen at fields corresponding to $\theta_B = (2n + 1) \pi /2$, for integer $n$. Here, $g$ is the Land{\'e} $g$-factor, $\mu_B$ the Bohr magneton, $L$ the junction length, and $v_F$ the Fermi velocity. For an InSb nanowire of length $500 - 1000\; \mathrm{nm}$, the first cusp ($\theta_B = \pi/2$) is estimated in Ref. \cite{supp_so_zeeman} to be at $B = 0.2\; \mathrm{T}$. However, using the values for our InAs nanowire device, which is shorter ($L \simeq 200\; \mathrm{nm}$), and has a $g$-factor that is roughly a factor of 5 smaller than that of InSb, we estimate $B \sim 7 \mathrm{T}$ for the position of the first cusp. This is an order of magnitude larger than the observed value of $B \sim 0.6\; \mathrm{T}$. Furthermore, the regularity of positions, in $B$, of $I_c$ minima predicted by the relation $\theta_B = (2n + 1) \pi /2$ does not hold for the InAs junction. In fact, we find the positions of the minima can be tuned using the voltages on the local gates, see Figure 3 in the main text. Also, the observed value of $I_c$ at the first antinode (e.g. at $B_\parallel \simeq 0.8$ T in Figure 3a) is roughly 0.15 times the value of $I_c$ at zero field, whereas in \cite{supp_so_zeeman} this ratio is predicted to be close to 1.\\
\indent Therefore, it is concluded that the $I_c$ vs $B_\parallel$ oscillations observed in the InAs nanowire junction are not due to the interplay of spin-orbit and Zeeman effects. We neglect these effects in our analyses for simplicity. Observing 0-$\pi$ transitions in an InAs nanowire junction may be difficult, because the high magnetic field required to drive the transition will likely destroy superconductivity in the leads.
\section{Magnetic Field Misalignment}
\label{sec:misalign}
\indent The perpendicular magnetic field, $B_\perp$, was applied in the plane of the device substrate ($xy$-plane in figure 1a, main text). The deviation angle of the field from the perpendicular ($\hat{\bo y}$) direction is $2^\circ \pm 3^\circ$. Since $\mathrm{tan}(2^\circ) \simeq 0.03$, this misalignment was neglected.\\
\indent The axial magnetic field, $B_\parallel$, was applied in the $xy$-plane, at an angle $\alpha = 8^\circ \pm 4^\circ$ with respect to the nanowire axis $\hat{ \bo {x}}$. The axial component of the field is $B_\parallel \times \mathrm{cos} \: \alpha$. Since $\mathrm{cos} \: \alpha > 0.98 \simeq 1$, the axial component is very close in magnitude to $B_\parallel$.\\
\indent The perpendicular ($\hat {\bo y}$) component of $B_\parallel$ is of magnitude $B_\parallel \times \mathrm{sin}\: \alpha$. This results in a quasi-Gaussian decay of the critical current $I_c$ with $B_\parallel$, due to a pair-breaking mechanism effectively equivalent to spin-flip scattering (see Section \ref{sec:perp}). Let us quantify this quasi-Gaussian decay by a suppression factor $s (B_\parallel) = I_c (B_\parallel) / I_c (B_\parallel=0)$. Given that $\mathrm{sin}\: \alpha$ is in the range $0.07 - 0.2$, with an average value $0.13$, we estimate $s$ is close to 1 ($s > 0.75$) for $B_\parallel < 1.5 \; \mathrm{T}$, covering a large range of the measured magnetic fields. However, for the highest field measured, $B_\parallel = 2.5 \; \mathrm{T}$, the $s$ can be as severe as $0.3$.\\
\indent A similar field dependence can be defined for the critical current density, by generalizing $J_c (\theta)$ to $J_c(B_\parallel, \theta) = s(B_\parallel)J_c(\theta)$. The $B_\parallel$- and $\theta$-dependent parts of $J_c$ are separable, and the pair breaking effect of $B_\parallel$ has been neglected in the treatment given in the main text (i.e. $s(B_\parallel) = 1$ for all $B_\parallel$) . Including the field dependence of $J_c$ will not qualitatively change the interference pattern (see also discussion in Section \ref{sec:interplay}). However, $I_c$ and a field-dependent $J_c$ will no longer be simply related as a Fourier pair. The broadening parameter $\sigma$, which influences the attenuation of $I_c$ at high fields (see Section \ref{sec:broadening}), is expected to be sensitive to the field dependence of $J_c$. We leave the treatment of a model with a field dependant critical current density to future work.
\section{The Effect of Broadening the Peaks in the Spectral Density}
\label{sec:broadening}
\indent In the quasi-ballistic regime, the width of the Gaussian peaks in the modelled critical current density, $J_c(\theta)$, is parametrized by $\sigma$ (see Eq. 3 in the main text). The broadening of the peaks determines the qualitative behaviour of $I_c (B_\parallel)$. To illustrate this we use the example developed in the main text, where $n_l$ = 1 for $l = -3, -2, \ldots, 3$, and zero otherwise. We show in Figure  \ref{fig:model_supp} the critical current density $J_c$ vs. the magnetic frequency  $\Phi_0 / \Phi_{\mathrm{nw}}$, and the resulting $I_c$ vs. $\Phi_{\mathrm{nw}} / \Phi_0$, for two different values of $\sigma$. When the spectral density $J_c$ shows sharp peaks in its distribution vs. winding angle $\theta$ (i.e., when $\sigma \ll \mathrm{d}\theta_l / \mathrm{d}l$), $I_c$ oscillates. However, when the distribution of $J_c$ is very broad ($\sigma \gg \mathrm{d}\theta_l / \mathrm{d}l$), $I_c$ shows a monotonic, quasi-Gaussian decay. The model was fit to the experimental data using an intermediate regime of broadening.
\begin{figure}[b!]
\stepcounter{newfigure}
\centering
\includegraphics[width=0.6\textwidth]{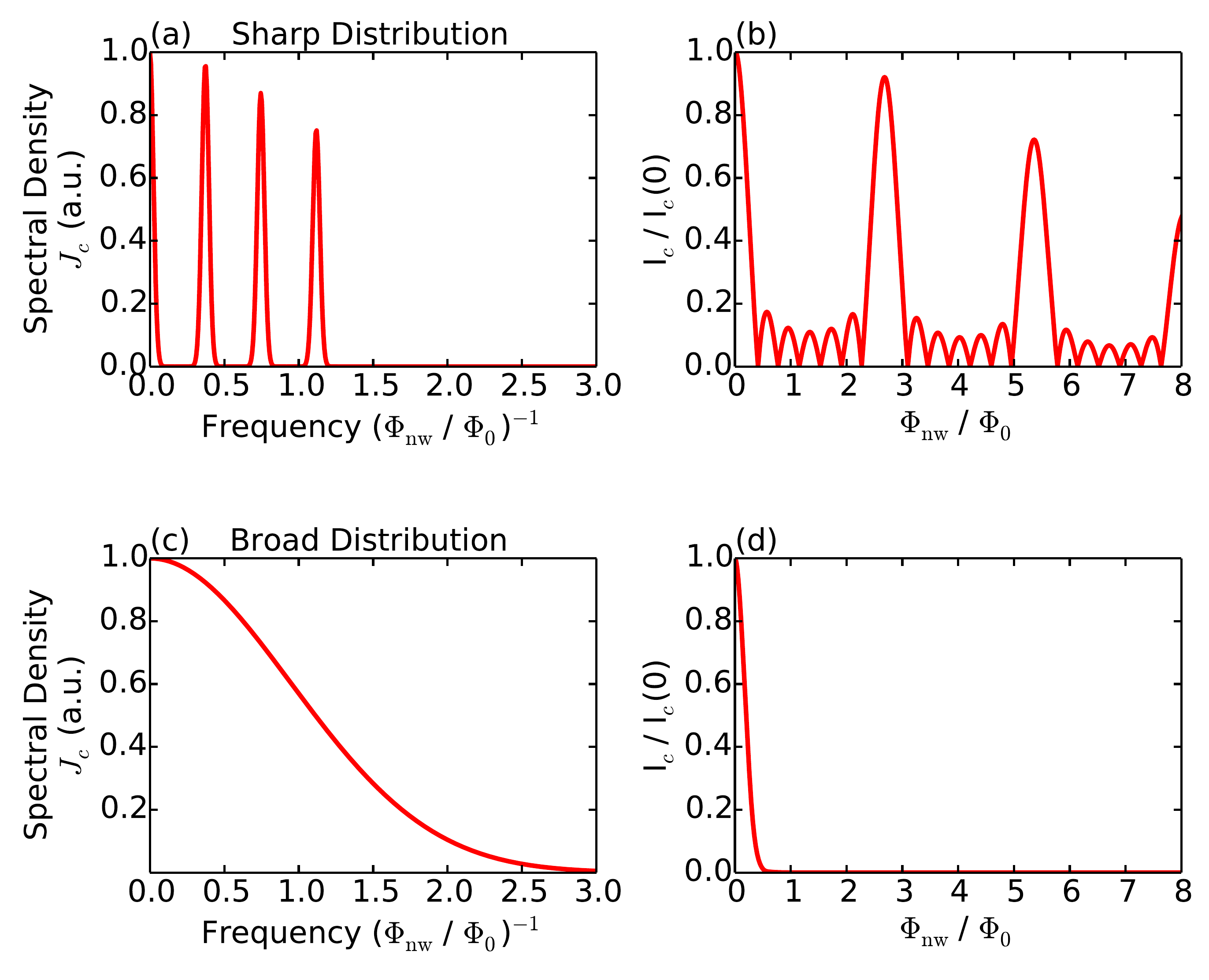}
\caption{\label{fig:model_supp}
Spectral densities (left) critical currents (right) in the quasiballistic regime, as a function of the broadening parameter $\sigma$. The spectral density $J_c$ is plotted vs. the magnetic frequency $(\Phi_0 / \Phi_{nw})$, see Figure 2 in the main text. The envelope function $J_{max} (\theta)$ has been taken into account, assuming $l_{in} = 400\; \mathrm{nm}$ as the inelastic scattering length (see Section \ref{sec:usadel}).
(a,b) $\sigma = 0.15$. Sharp peaks in $J_c$ compared to the peak spacing, $\sigma \ll \mathrm{d}\theta_l / \mathrm{d}l$, results in an oscillatory $I_c$. The oscillations in (b) attenuate with increasing magnetic flux, due to the finite $\sigma$. Panels (a,b) are reproduced in the bottom row of Figure 2 in the main text.
(c,d) $\sigma = 5.0$. Broad peaks in $J_c$ compared to the peak spacing, $\sigma \gg \mathrm{d}\theta_l / \mathrm{d}l$. The peaks in (c) overlap, creating a bell shaped curve. The attenuation of critical current with increasing magnetic flux is strong enough that no oscillation are seen in (d), resulting in a monotonic, quasi-Gaussian decay.
The following parameters (defined in the main text) were used in these examples: $d = 63\; \mathrm {nm}$, $L = 200\; \mathrm{nm}$, $l_e = 80\; \mathrm{nm}$, $E_F = 150\; \mathrm{meV}$. $n_l = 1$ for $l = -3, -2, \ldots, 3$, and zero otherwise. The experimental data (Figure 3 in the main text) is best fit to an intermediate regime, with $\sigma \sim 0.9$.}
\end{figure}\\
\section{Varying the Nanowire Diameter}
\label{sec:diameter}
We discuss the effect of changing the nanowire diameter $d$ on the interference effect in the quasi-ballistic model, where all carriers are on a cylindrical shell at a radius $d/2$ from the nanowire center. Consider the critical current of the junction as a function of the axial magnetic field, $B_\parallel$. Eq. 2 in the main text reads
\begin{equation}
I_c (B_\parallel) =  \left| \int_{-\infty}^\infty J_c(\theta) \mathrm{e}^ {\left({ i \frac{ \Phi_{\mathrm{nw}} }{ \Phi_0 } \theta} \right)} \mathrm d\, \theta \right|,
\label{Eq_d_Ic}
\end{equation}
with $\Phi_{nw} = \pi d^2 B_\parallel / 4$. The critical current density, $J_c$, is given by Eq. 3 in the main text as
\begin{equation}
J_c (\theta) = J_{\mathrm{max}} (\theta) \sum_l \frac {n_l}{\sigma \sqrt{2 \pi}} \mathrm{exp} \left( - \frac{ (\theta - \theta_l)^2} {2 \sigma ^2} \right).
\label{Eq:crit_curr_density_in_sec_d}
\end{equation}
\indent Let us change the diameter from $d$ to a value $d^\prime$, resulting in  $\Phi_{nw} \rightarrow \Phi^\prime_{nw} = (\frac{d^{\prime2}}{d^2}) \Phi_{nw}$, and $J_c (\theta) \rightarrow J_c^\prime (\theta)$, $I_c (B_\parallel) \rightarrow I_c^\prime (B_\parallel).$ We calculate $J_c^\prime (\theta)$ below, and show that $I_c^\prime (B_\parallel) = I_c(B_\parallel)$, up to a rescaling of the the broadening parameter $\sigma$, and a rescaling of the envelope function $J_{max} (\theta)$.\\
\indent We make the substitution of variables $\theta = (d^2 / d^{\prime 2}) \theta^\prime$ in Eq. \ref{Eq_d_Ic} to write
\begin{align}
I_c (B_\parallel) \rightarrow I_c^\prime (B_\parallel) & = \left| \int_{-\infty}^\infty J_c^\prime(\theta) \mathrm{e}^ {\left({ i \frac{ \Phi^\prime_{\mathrm{nw}} }{ \Phi_0 } \theta} \right)} \mathrm d\, \theta \right| \nonumber \\ 
\; & = \frac{d^2}{d^{\prime 2}}\left| \int_{-\infty}^\infty J_c^\prime \left( \left(\frac{d^2}{d^{\prime ^2}} \right) \theta^\prime \right) \mathrm{e}^ {\left({ i \frac{ \Phi_{\mathrm{nw}} }{ \Phi_0 } \theta^\prime} \right)} \mathrm d\, \theta^\prime \right|.
\end{align}
In the above integral, $\theta, \theta^\prime$ are dummy variables, and it suffices to have $J_c^\prime \left( \theta\right)  = (d^{\prime 2} / d^2) J_c (\theta^\prime)$ in order to get $I_c^\prime (B_\parallel) = I_c(B_\parallel)$. Let us use a constant envelope function $J_{max} (\theta) = J_0$ for now. Notice that in Eq. 5 in the main text, $v_\theta \propto 1/d$, so the positions of the peaks, $\theta_l$ have an inverse-square dependence od $d$, so when $d \rightarrow d^\prime$, $\theta_l \rightarrow \theta_l^\prime = (d^2/d^{\prime 2}) \theta_l$. Using Eq. \ref{Eq:crit_curr_density_in_sec_d} we write
\begin{align}
J_c^\prime \left(\left(\frac{d^2}{d^{\prime 2}}\right)\theta^\prime \right) &=  J_0 \sum_l \frac {n_l}{\sigma \sqrt{2 \pi}} \mathrm{exp} \left( - \frac{ \left( (d^2/d^{\prime 2})\theta^\prime  - \theta_l^\prime \right)^2} {2 \sigma ^2} \right) \nonumber \\ 
\; &= J_0 \sum_l \frac {n_l}{\sigma \sqrt{2 \pi}} \mathrm{exp} \left( - \frac{ \left(d^2/d^{\prime 2} \right)^2  \left( \theta^\prime - \theta_l \right)^2} {2 \sigma ^2} \right).
\end{align}
If we rescale $\sigma$ as $\sigma \rightarrow \sigma^\prime = (d^2 / d^{\prime 2}) \sigma$, the factors in the exponent cancel, and we get the result $J_c^\prime \left( (d^2/d^{\prime 2})\theta^\prime \right)  = (d^{\prime 2} / d^2) J_c (\theta^\prime)$, or equivalently $I_c^\prime(B_\parallel) = I_c (B_\parallel)$.\\
\indent Let us consider the case of a generic $J_{max} (\theta)$. Using similar analysis we see that, if the angle argument of $J_{max}$ is rescaled such that $J_{max}(\theta) \rightarrow J^\prime_{max} (\theta) = J_{max}( (d^{\prime 2}/d^2) \theta)$, then the relation $J_c^\prime \left( (d^2/d^{\prime 2})\theta^\prime \right)  = (d^{\prime 2} / d^2) J_c (\theta^\prime)$ holds true (up to rescaling $\sigma$) . This follows from the dependence of $J_{max}$ on the length of the spiral paths $l_\theta = \sqrt{L^2 + (d \theta/2)^2}$, as discussed in Section \ref{sec:usadel}. When the diameter of the nanowire becomes very large, the spiral paths on its circumference become very long, and the supercurrent is suppressed due to inelastic scattering.\\
\indent In summary, we have shown that if the radial position of the carriers is changed, the spectral density $J_c (\theta)$ is unaffected up to a rescaling of the envelope function ($J_{max}$) and the width of its peaks ($\sigma$). Crucially, the frequencies at which the peaks of $J_c$ appear, i.e. the frequencies of oscillations of $I_c$ vs. $B_\parallel$, are unaffected. We suggest that, for general radial wavefunctions of the carriers, a Josephson interference effect similar to that discussed in the main text, and with the same periods of oscillations in $B_\parallel$, should appear. However, a detailed calculation is required to verify this, using Usadel equations and taking into account the complete Hamiltonian for the system, including the radial confinement potential.

\section{$J_{max}$ calculated by the Usadel Equations}
\label{sec:usadel}
We use the quasi-classical Green's functions theory \cite{supp_Belzig_quasiclassical} to describe the proximity effect superconductivity in the InAs nanowire junction. Further details on this approach can be found in \cite{supp_gueron}, on which this section is based.\\
\indent The starting point is a field operator 
\[
\Psi = \left( \begin{array}{c} \Psi_\uparrow (x,t) \\ \Psi^\dagger_\downarrow (x,t) \end{array} \right)
\]
acting on the electron-hole Nambu space. $\Psi_\uparrow (x,t)$ and $\Psi^\dagger_\uparrow (x,t)$ are annihilation and creation operators for a fermionic quasiparticle with spin $\uparrow$ at position $x$ and time $t$.\\
\indent The basic objects in terms of which the theory is developed are the Retarded, Advanced, and Keldysh Green's functions  ($\hat R, \hat A$, and $\hat K$, respectively) defined in terms of the field operator $\Psi$. We consider the nanowire to be in thermal equilibrium, so  the function $\hat K$ is redundant and holds no further information about the system than $\hat R, \hat A$. Being metallic superconductors, the Nb leads exhibit electron-hole symmetry. We assume electron-hole symmetry in the normal section as well, so $\hat A$ is also redundant. We concentrate on $\hat R$.
Using the standard angular ($\Theta,\phi$) parametrization on the unit sphere, we write
\begin{equation*}
\hat R = \mathrm{cos} \Theta \tau_z + \mathrm{sin} \Theta (\mathrm{cos} \phi \tau_x + \mathrm{sin} \phi \tau_y) = \left( \begin{array}{cc}
\mathrm{cos} \Theta & e^{-i \phi} \mathrm{sin} \Theta \\
e^{i \phi} \mathrm{sin} \Theta & - \mathrm{cos} \Theta
\end{array}
\right), \label{Eq1_retarded_GF}
\end{equation*}
where $\tau_{x,y,z}$ are Nambu spinors acting on the electron-hole degree of freedom. Here, $\Theta = \Theta(x,E)$ is the complex pairing angle which quantifies the strength of superconducting-like correlations (off-diagonal elements of $\hat{R}$) and normal-like correlations (diagonal elements of $\hat{R}$). We use the capital symbol $\Theta$ instead of the more standard $\theta$ to avoid confusion with winding angles. $\phi = \phi(x,E)$ is the real superconducting phase. We have considered only one spatial dimension $x$, along the axis of the nanowire. This one dimensional (1D) formulation is in anticipation of reducing the description of the nanowire to a quasi-1D model, see below.\\
\indent Our goal here is to calculate $\Theta (x,E)$ and $\phi (x,E)$ for all positions and energies, for the geometry of the junction. The equilibrium Usadel equations, governing $\Theta, \phi$, can be derived from the equation for $\hat{R}$. The 1D set of coupled equations reads:
\begin{subequations}
\begin{align}
\frac{\hbar D}{2} \frac{\partial^2 \Theta}{\partial x^2} + \left( iE - \frac{\hbar}{\tau_{in}}- \left( \frac{\hbar}{\tau_{sf}} + \frac{\hbar D}{2} \left( \frac{\partial \phi}{\partial x} + \frac{2e}{\hbar}A_x \right)^2 \right)  \mathrm{cos}\Theta  \right) \mathrm{sin}\Theta + \Delta(x)\mathrm{cos}\Theta = &\; 0,\\
\frac{\partial}{\partial x} \left( \left(\frac{\partial \phi}{\partial x} + \frac{2e}{\hbar} A_x\right) \mathrm{sin}^2 \Theta \right) = &\; 0.
\end{align} \label{Eq2_Usadel_1d}
\end{subequations}
Here, $D = l_e v_F/3$ is the diffusion coefficient of electrons, $l_e$ is the elastic mean free path and $v_F$ the Fermi velocity. The timescales for inelastic and spin-flip scatterings are denoted by $\tau_{in}$ and $\tau_{sf}$, respectively. $A_x$ is the axial component of the magnetic vector potential, due to a perpendicular external magnetic field, $B_\perp$. Spin-flip scattering off of magnetic impurities is ignored, so the only contribution to the spin-flip rate is the narrow-junction limit depairing term $\hbar / \tau_{sf} = e^2 d^2 D B_\perp^2 / (6 \hbar)$, as discussed in \cite{supp_Hammer_non-ideal-interface}. The quasi-1D depairing term is valid because the width of the junction is on the same order as the superconducting coherence length in Nb, $d = 63\; \mathrm{nm} \sim \xi_{Nb}$. The parallel component of the magnetic field does not enter the 1D Usadel equations. We use $\Delta$ for the superconducting energy gap due to electron-phonon coupling. In the Nb leads, this equals $\Delta_{Nb} = 1.2\; \mathrm{meV}$, and in the normal section of the junction we have $\Delta = 0$.\\
\indent All physical quantities related to the junction can be derived from $\Theta$ and $\phi$. In particular, the supercurrent density in the N (S) section of the junction is
\begin{equation}
J_{N(S)} = -(\sigma_{N(S)} / e) \int^\infty _0 \mathrm{d} E\; \mathrm{tanh} \left( E / 2k_B T \right) \mathrm{Im} \left( \mathrm{sin} ^2 \Theta \right) \left( \frac{\partial \phi}{\partial x} + (2e / \hbar) A_x \right).
\label{Eq3_supercurrent_density}
\end{equation}
Here, $\sigma_{N(S)}$ is the normal state conductivity of the normal (superconducting) section of the junction, $k_B$ the Boltzmann constant and $T$ the temperature. By comparison with the Ginzburg-Landau (GL) result $\mathbf{J} = (-\hbar e / m^*) |\psi_{GL}|^2 (\bo {\nabla} \phi + (2e / \hbar) \mathbf{A})$, we identify the modulus squared of the GL order parameter in the normal section as
\begin{equation}
| \psi_{GL} |^2 = (m^* \sigma_{N} / e^2 \hbar) \int^\infty _0 \mathrm{d} E\; \mathrm{tanh} \left( E / 2k_B T \right) \mathrm{Im} \left( \mathrm{sin} ^2 \Theta \right),
\label{Eq4_GL_psi2}
\end{equation}
where $m^*$ is the effective mass. The energy dependence of the phase $\phi$ has been neglected. This is justified because the Nb reservoirs at the ends of the junction are bulk superconductors, with an energy independent phase difference $\gamma$ determined by the bias current. The N-S interface are transparent with $t \sim 0.65$, so the phase gradient in the normal section of the junction is also energy independent \cite{supp_anne_phd}.\\
\indent In the model for the InAs nanowire junction, orbital effects are the dominant mechanism for Josephson interference. However, the field operator $\Psi$ contains no information on the orbital structure of the nanowire, such as angular momentum subbands -- in fact, it acts to create/annihilate a fermion of angular momentum zero. This is because the effects of quantum confinement are neglected in the standard treatment of Green's functions that leads to Eq. \ref{Eq2_Usadel_1d}, e.g. the treatment given in \cite{supp_gueron}. It is possible, in principle, to generalize the Green's functions method by starting from a Hamiltonian for the junction which contains the quantum confinement potential, then deriving the orbital subband structure of the nanowire, and the effects of its interplay with superconductivity. However, such treatment is beyond the scope of this manuscript. Instead, in the main text we assume a sinusoidal current-phase relationship and model the orbital effects using a semi-classical approach based on the spiral trajectories of particles. What we wish to calculate here is the suppression of the supercurrent, for long trajectories, due to inelastic scattering.\\
\indent Consider a spiral trajectory on the circumference of the nanowire with winding angle $\theta$. The length of this trajectory is $\l_\theta = \sqrt{L^2 + (d \theta /2)^2}$, where $L = 200\; \mathrm{nm}$ is the length of the junction and $d = 63\; \mathrm{nm}$ the diameter of the nanowire. We approximate the supercurrent density of this trajectory by that of a planar junction (Eq. \ref{Eq3_supercurrent_density}) of length $L^\prime = l_\theta$. In doing so we are modelling the nanowire as a set of parallel, narrow (quasi-1D) planar junctions, each with a length $L^\prime = l_\theta$, corresponding to a winding angle $\theta$. What limits the supercurrent in the planar junction is the magnitude squared of the GL order parameter at the ``bottleneck'' in the normal section midway between the Nb leads. Since $|\psi_{GL}|^2$ does not depend on the phase, we set $\phi$ equal to zero everywhere, and the Usadel equations (Eq. \ref{Eq2_Usadel_1d}) simplify to
\begin{equation}
\frac{\hbar D}{2} \frac{\partial^2 \Theta}{\partial x^2} + \left( iE - \frac{\hbar}{\tau_{in}} \right)\mathrm{sin}\Theta + \Delta(x) \mathrm{sin}\Theta = 0.
\label{Eq5_simple_usadel}
\end{equation}
Here we have set the spin-flip scattering rate $\tau_{sf}$ equal to zero, i.e. no external magnetic field applied. We discuss this assumption below. Let the N-S boundaries be at $x = 0, x = L^\prime$. We have
\begin{equation*}
\Delta(x) = 
\begin{cases}
\Delta_{Nb}  =1.2\; \mathrm{meV} & \quad x < 0\; \text{or}\; x > L^\prime \\
0 & \quad \text{otherwise}
\end{cases}
\end{equation*}
The pairing angle in the superconducting Nb leads, $\Theta_S$, tends to its BCS value for Nb at positions far from the N-S interfaces, $\Theta_S = \Theta_{BCS}$, for $x \rightarrow \pm \infty$. Here,
\begin{equation*}
\Theta_{BCS} (E) =
\begin{cases}
\pi/2 + i\; \text{argtanh} (E/\Delta_{Nb}) & \quad \text{if} \; |E| < \Delta_{Nb}, \\
i\; \text{argtanh} (\Delta_{Nb}/E) & \quad \text{if} \; |E| > \Delta_{Nb}.
\end{cases}
\end{equation*}
At the N-S interfaces, the pairing angle in the normal section, $\Theta_N$, and the superconducting section, $\Theta_S$, are subject to the following continuity condition:
\begin{equation*}
\left. \sigma_N \frac{\partial \Theta_N }{\partial x} \right|_{x =x_i} = \left. \sigma_S \frac{\partial \Theta_S }{\partial x} \right|_{x =x_i} = g_{b}\; \mathrm{sin} \left( \Theta_S (x_i,E) - \Theta_N (x_i, E) \right),
\end{equation*}
for $x_i = 0, L^\prime$. This expresses the conservation of spectral current at the boundaries. Here, $g_b = G_b / \mathcal{A}_b$ is the conductance of an interface normalized by its area. We can further simplify the situation by noticing the high conductance at the interfaces, $r = R_b / R_N = G_N / G_b \sim 0.15 \ll 1$, see Section \ref{subsec:contactres}. So, we use the transparent interface limit \cite{supp_golubov} to write
\begin{equation}
\Theta_S (x = x_i) = \Theta_N (x = x_i) = \Theta_{BCS},\quad \text{for } x_i = 0,L^\prime.
\label{Eq6_boundary_cond}
\end{equation}
\begin{figure}[t!]
\stepcounter{newfigure}
\centering
\includegraphics[width=300px]{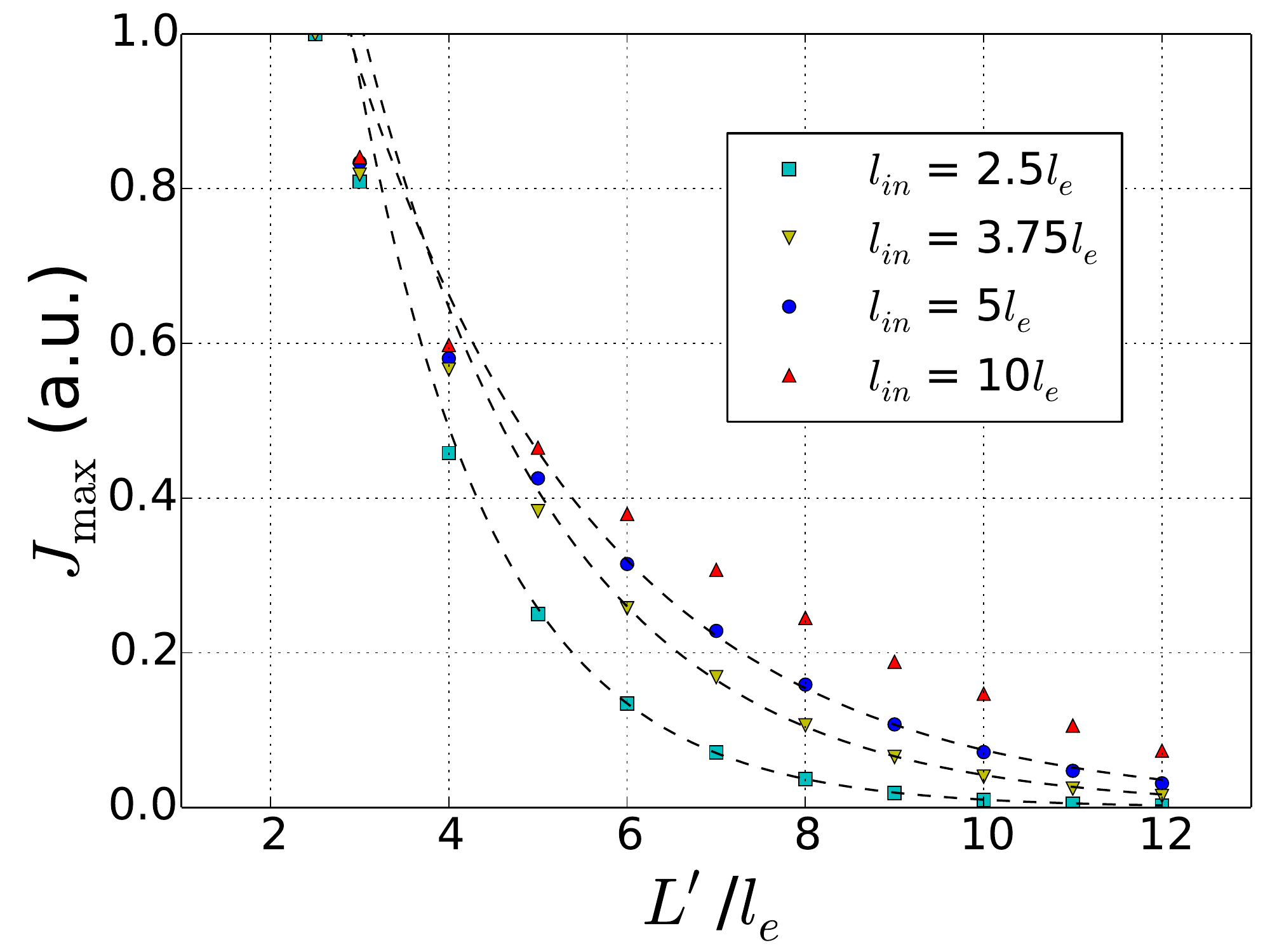}
\caption{\label{fig:usadel}
The envelope function $J_{max}$ (Eq. \ref{Eq7_jmax}) vs the normalized planar junction length $L^\prime / l_e$, as a function of the inelastic scattering length $l_{in}$. Here, $l_e = 80\; \mathrm{nm}$ is the elastic scattering length. The supercurrent density decays as the length of the trajectory is increased, and the decay is faster for shorter $l_{in}$. The dashed lines are exponential fits to the region $L^\prime / l_e > 5$, for the cases $l_{in}/l_e = 2.5,\; 3.75,\; 5$. Each curve is normalized to its value at $L^\prime = 200\; \mathrm{nm}$ (i.e. $L^\prime / l_e = 2.5$). For a trajectory with winding angle $\theta$ on the circumference of the nanowire, $J_{max} (\theta)$ can be calculated by setting $L^\prime = l_\theta = \sqrt{L^2 + (d \theta /2)^2}$. The curve corresponding to $l_{in} = 5l_e$ is used in the inset of Figure 2 in the main text.}
\end{figure}
\indent We solve Eq. \ref{Eq5_simple_usadel} for the complex quantity $\Theta_N$ numerically, subject to the boundary conditions in Eq. \ref{Eq6_boundary_cond}. We insert the solution $\Theta_N$ into Eq. \ref{Eq4_GL_psi2} to calculate $|\psi_{GL}|^2$ at middle of the junction, i.e. $x=L^\prime/2$. We do this calculation for different values of the junction length $L^\prime$.   The decay of the supercurrent with trajectory length $l_\theta$ is captured by the envelope function
\begin{equation}
 J_{max} (\theta) = \frac{ \left| \psi_{GL} \left(L^\prime = l_\theta , x = L^\prime /2 \right) \right|^2}{ \left| \psi_{GL} \left(L^\prime = L , x = L^\prime /2 \right) \right|^2} ,
\label{Eq7_jmax}
\end{equation}
where $L = 200\; \mathrm{nm}$ is the length of the shortest path across the junction, with $\theta = 0$, i.e. a  straight trajectory.\\
\indent In Figure \ref{fig:usadel} we show $J_{max}$ vs $L^\prime / l_e$ as a function of the inelastic scattering length $l_{in}$. As expected, $J_{max}$ decays as the length of the trajectory is increased, and the decay is faster for shorter $l_{in}$. For $L^\prime$ much longer than $l_e$, and longer than $l_{in}$, the $J_{max}$ curves can be fitted to exponentials. The length scale of the exponential decay depends on $l_e$ as well as $l_{in}$. For short lengths, $L^\prime \lesssim l_e$, the $J_{max}$ curves saturate (not shown). The parameters used for this calculation are the same as those used in figure 2 in the main text: $d = 63\; \mathrm {nm}$, $L = 200\; \mathrm{nm}$, $l_e = 80\; \mathrm{nm}$, $l_{in} = 400\; \mathrm{nm}$, $E_F = 150\; \mathrm{meV}$, $m^* = 0.023 m_e$, and $T = 100\; \mathrm{mK}$ is taken as the (electron) temperature.\\
\indent Finally, we comment on setting the spin-flip scattering rate $\Gamma_{sf} = \hbar / \tau_{sf}$ equal to zero in the above calculation. The axial magnetic field $B_\parallel$ can create a pair-breaking mechanism (i.e. an effective spin-flip mechanism similar to that in Section \ref{sec:perp}) for Andreev pairs on spiral paths, because the azimuthal ($\hat{\bo \theta}$) component of the their velocity is perpendicular to $B_\parallel$. However, this effect is overshadowed by the pair-breaking due to the misalignment of $B_\parallel$ with respect to the nanowire axis, as discussed in Section \ref{sec:misalign}. This is because the $\hat{\bo \theta}$ component of the pair velocity is small compared to its total (Fermi) velocity between scattering events: $v_\theta \lesssim 0.3 v_F$. A typical value is $v_\theta \sim 0.1 v_F$, for a subband with angular momentum quantum number $l = 1$. Furthermore, the axial cross section of the nanowire, $\pi d^2 / 4$, is a factor of 4 smaller than its perpendicular cross section, $d \times L$. Estimates give $J_{max} (\Gamma_{sf}) / J_{max} (\Gamma_{sf} = 0) \gtrsim 0.8$ for the largest magnetic field measured, $B_\parallel = 2.5\; \mathrm{T}$. As in Section \ref{sec:misalign}, we neglect such field dependence of $J_c$.

%\bibliographystyle{h-physrev}
%\bibliography{BIBsupp.bib}
%\putbib[BIBsupp]
%\end{bibunit}
\end{widetext}
\end{document}